\documentclass[utf8]{frontiersSCNS} 

\usepackage{url,hyperref,lineno,microtype,subcaption}
\usepackage{amsmath}
\usepackage{amsfonts}

\def\keyFont{\fontsize{8}{11}\helveticabold }
\def\firstAuthorLast{Bl\'azquez-Salcedo {et~al.}} 
\def\Authors{Jose Luis Bl\'azquez-Salcedo\,$^{1,*}$, Fech Scen Khoo\,$^{1,2}$, Jutta Kunz\,$^{2}$
and Vincent Preut\,$^{2}$}
\def\Address{$^{1}$Departamento de F\'isica Te\'orica II and IPARCOS, Facultad de Ciencias F\'isicas, 
Universidad Complutense de Madrid, 28040 Madrid, Spain\\
$^{2}$Institute of Physics, University of Oldenburg,
	26111 Oldenburg, Germany}
\def\corrAuthor{Jose Luis Bl\'azquez-Salcedo}

\def\corrEmail{jlblaz01@ucm.es}

\begin{document}
\onecolumn
\firstpage{1}

\title[Polar Quasinormal Modes]{Polar Quasinormal Modes of Neutron Stars in Massive Scalar-Tensor Theories } 

\author[\firstAuthorLast ]{\Authors} 
\address{} 
\correspondance{} 

\extraAuth{}

\maketitle

\begin{abstract}
We study polar quasinormal modes of relativistic stars in
scalar-tensor theories,
where we include a massive gravitational scalar field
and employ the standard Brans-Dicke coupling function.
For the potential of the scalar field we consider a simple mass
term as well as a potential associated with $R^2$ gravity.
The presence of the scalar field makes the spectrum of quasinormal
modes much richer than the spectrum in General Relativity.
We here investigate radial modes ($l=0$) and quadrupole modes
($l=2$). The general relativistic $l=0$ normal modes
turn into quasinormal modes in scalar-tensor theories, 
that are able to propagate outside of the stars.
In addition to the pressure-led modes 
new scalar-led $\phi$-modes arise.
We analyze the dependence of the quasinormal mode frequencies
and decay times on the scalar field mass.

\tiny
 \keyFont{ \section{Keywords:} Relativistic stars: structure, stability, and oscillations, Modified theories of gravity, Gravitational waves} 
\end{abstract}


\section{Introduction}

Following the first direct detections of 
gravitational waves, mostly emitted from merging black holes \citep{LIGOScientific:2016aoc,LIGOScientific:2016sjg, LIGOScientific:2017ycc,LIGOScientific:2017bnn,LIGOScientific:2018mvr},
there have been numerous further detections. 
Detections of merging neutron stars are much rarer events
\citep{LIGOScientific:2017zlf,LIGOScientific:2020aai}, 
including, in particular, the observation of the electromagnetic
counterpart, which has opened the age of multi-messenger
gravitational wave astronomy 
\citep{Coulter:2017wya, TheLIGOScientific:2017qsa, GBM:2017lvd,Abbott:2018wiz}. 
With this new channel of observation, it is possible to directly test the strong gravity regime. 
Therefore, the study of the properties of the gravitational waves emitted by astrophysical sources 
has become ever more important.  

Gravitational waves from merging compact objects possess three phases, inspiral, merger and ringdown.
The resonant frequencies that dominate the ringdown can be studied using quasinormal modes (QNMs) 
\citep{Kokkotas:1999bd,Nollert:1999ji,Berti:2009kk,Konoplya:2011qq}. 
Whereas currently the precision of the detected signals is mostly considered too low
to extract information from the ringdown, detectors are expected to achieve the necessary sensitivity 
in the near future \citep{Berti:2018vdi,Barack:2018yly}.
However, already now first attempts have been made
to extract not only the dominant QNM but also subdominant QNMs from the ringdown phase
\citep{Giesler:2019uxc,Bhagwat:2019bwv,JimenezForteza:2020cve,Capano:2021etf}.
Clearly, QNMs represent crucial observables.

In the case of the QNMs of neutron stars, one additional difficulty is that their properties
including their QNM spectrum depend on the equation of state (EOS),
which describes the effective relation between energy density and pressure of the matter 
that composes the interior of the star. So far, the EOS is not well understood, 
and numerous models have been proposed (see e.g., \citep{Haensel:2007yy}). 
Nonetheless, current observations allow to impose many constraints on the EOS  
\citep{Lattimer:2013hma,Antoniadis:2013pzd,Ozel:2016oaf,Most:2018hfd}
and since the spectrum of neutron stars is much richer than the spectrum of black holes, 
even constraints on the theory of gravity are possible \citep{Berti:2018vdi,Berti:2018cxi, Berti:2015itd}. 
Moreover, \textit{universal relations} between properly scaled global parameters of the neutron stars 
provide almost EOS independent relations, which can strongly enhance the analysis of neutron stars
\citep{Yagi:2016bkt,Doneva:2017jop}.

The spectrum of QNMs of neutron stars has already been well studied in General Relativity 
\citep{Andersson:1996pn,Andersson:1997rn,Kokkotas:1999mn,Benhar:2004xg,BlazquezSalcedo:2012pd,Blazquez-Salcedo:2013jka,Mena-Fernandez:2019irg,Volkel:2019gpq}. 
When the background spacetime is static and spherically symmetric, 
the perturbations decouple into two independent channels, the axial (odd-parity) channel 
and the polar (even-parity) channel. 
Axial perturbations couple only to spacetime oscillations, 
and possess the so called rapidly damped w-modes \citep{Kokkotas:2003mh}. 
Polar perturbations couple also to the matter of the star, thus their spectrum is much richer. 
In the simplest case one finds pressure-driven modes (the fundamental f-mode and the excited p-modes) 
as well as spacetime modes (w-modes).

A neutron star may possess, however, still further modes.
In General Relativity these are a set of 
radial \textit{normal modes}. These represent undamped modes, 
that turn into unstable modes when for a given equation of state
the maximum of the neutron star mass is passed
\citep{Chandrasekhar:1964zz,Chandrasekhar:1964zza,1966ApJ...145..505B,1966ApJ...145..514M,1977ApJ...217..799C,1983ApJS...53...93G,1992A&A...260..250V,1998IJMPD...7...49D}. 
The usual nomenclature for these modes is the
fundamental F-mode and the excited H-modes.
In General Relativity these modes are irrelevant for the ringdown, though,
since radial perturbations cannot propagate 
outside the neutron stars.

Because of their extreme compactness
neutron stars represent excellent laboratories 
to test General Relativity
and alternative theories of gravity
(see e.g., \citep{Will:2005va,Faraoni:2010pgm,Berti:2015itd,CANTATA:2021ktz}).
For instance, \textit{universal relations} 
in alternative theories of gravity
can differ
appreciably from those of General Relativity
 \citep{Yagi:2016bkt,Doneva:2017jop}.
Moreover, the spectrum of neutron stars is typically
much richer in alternative theories of gravity,
since additional degrees of freedom are present.
Therefore 
constraints on alternative theories of gravity are possible \citep{Berti:2018vdi,Berti:2018cxi, Berti:2015itd}. 

We now focus on QNMs in alternative theories of gravity
with an additional scalar degree of freedom
(see e.g.~\citep{Blazquez-Salcedo:2018pxo}).
The QNMs of neutron stars in scalar-tensor theories
STTs, were first considered in 
\citep{Sotani:2004rq}.
Here the polar f- and p-modes were obtained
by making use of the Cowling approximation,
which simplifies the calculations considerably,
since the perturbations of the spacetime 
and the scalar field are frozen
(see also \citep{Yazadjiev:2017vpg} for the
inclusion of rotation).
The gravitational axial w-modes were studied first in
\citep{Sotani:2005qx}. 
As noted above, there is no coupling to the fluid or 
to the scalar field in the axial case,
making these studies much simpler than the polar ones.
For the axial QNMs the 
\textit{universal relations} were investigated 
\citep{AltahaMotahar:2018djk},
also for a massive scalar field with self-interaction
\citep{AltahaMotahar:2019ekm}.

When one ventures beyond the Cowling approximation
(see \citep{Blazquez-Salcedo:2020ibb,Kruger:2021yay} for its effects on the fundamental quadrupole mode),
qualitatively new types of polar modes arise,
since scalar radiation can be produced.
The detection of scalar radiation
would represent a most important discovery,
while the non-detection of scalar radiation
would allow to put constraints on the 
theory.
So far, the presence of scalar modes in STTs
has only been studied in an exploratory way.
The sector of radial neutron star oscillations
has been explored first in \citep{Mendes:2018qwo},
where scalar radiation from spontaneously scalarized 
neutron stars was addressed. 
Recently, some scalar modes and also quadrupole modes
have been obtained in massive Brans-Dicke theories \citep{Blazquez-Salcedo:2020ibb}.

Such massive Brans-Dicke theories
are closely related to $f(R)$ theories,
since the latter can be reformulated in terms of STTs
\citep{Sotiriou:2008rp,DeFelice:2010aj,Capozziello:2011et}.
Among these theories, in particular, $R^2$ gravity has
received much attention in recent years.
$R^2$ gravity is based on the Lagrangian $f(R) = R + a R^2$,
with coupling parameter $a$.
When reformulated and studied in the Einstein frame,
a scalar field potential results, 
where the mass for the scalar field depends on
the coupling parameter $a$,
$m_\varphi \sim a^{-1/2}$
\citep{Yazadjiev:2014cza,Staykov:2014mwa,Yazadjiev:2015zia,Astashenok:2017dpo}.
For $a\to 0$ the general relativistic limit is obtained, 
in contrast, for 
$a \to \infty$ a particular Brans-Dicke theory arises. 

Neutron star properties have been studied in $R^2$ gravity 
in \citep{Orellana:2013gn,Staykov:2014mwa,Yazadjiev:2014cza,Yazadjiev:2015zia,Astashenok:2017dpo}.
Axial QNMs have been obtained in
\citep{Blazquez-Salcedo:2018qyy,Blazquez-Salcedo:2018pxo}.
In $R^2$ gravity
the frequencies typically deviate significantly 
from the values in General Relativity,
but the damping times differ appreciably
only for small coupling constant $a$.
Polar modes have been studied in the Cowling approximation in $R^2$ theory \citep{Staykov:2015cfa}). 
Only recently, the full polar QNMs have been 
addressed without making
use of the Cowling approximation,
where, in particular, ultra long lived modes 
were shown to exist in the radial sector 
\citep{Blazquez-Salcedo:2020ibb}. 

The present paper is devoted to a more detailed study of
the QNMs of the full polar perturbations of realistic neutron stars in massive STTs, 
extending the previous analyses
\citep{Staykov:2015cfa,Blazquez-Salcedo:2018qyy,Blazquez-Salcedo:2020ibb}. 
The paper is organized as follows. 
In section 2 we revisit the static neutron stars
to fix the notation. 
In section 3 we provide the full polar perturbations for
neutron stars, and derive the equations and boundary
conditions describing the oscillations. 
In section 4 we make use of this formalism in order to
calculate the polar modes. 
We here focus on the $l=2$ fundamental mode 
and the $l=0$ radial modes, 
analyzing their dependence on the parameters
of the STTs, and the total mass of the configurations. 
We end the paper with our conclusions and an outlook.

\section{Static neutron stars in massive 
scalar-tensor theory} \label{sec_theory}

\vspace{0.5cm}

\subsection{Theoretical framework}

We consider massive STTs described by the action 
in the Einstein frame ($G=c=1$) \citep{Wagoner:1970vr}
\begin{equation}
S [g_{\mu\nu},\phi] = \frac{1}{16\pi } \int d^4x \sqrt{-g} 
\big( R - 2\partial_{\mu}\phi \, \partial^{\mu}\phi 
- V(\phi) + L_{M}(A^2(\phi)g_{\mu\nu},\chi) \big)~,
\label{EinsteinAction}
\end{equation}
with the metric $g_{\mu\nu}$, the curvature scalar $R$,
the scalar field $\phi$, the scalar potential $V(\phi)$,
the nuclear matter action $L_M$, and 
the standard Brans-Dicke coupling function
\begin{equation}
    A(\phi)= e^{-\frac{1}{\sqrt{3}}\phi} \ .
\end{equation}

We choose two examples for the potential $V(\phi)$ \citep{Faraoni:2009km}: 
$V_I$ represents simply a mass term, 
and $V_{II}$ is related to $R^2$ gravity
\citep{Faraoni:1999hp,Yazadjiev:2014cza,Yazadjiev:2015zia,Bhattacharya:2017pqc}
\begin{eqnarray}
V_I &=& 2 m_{\phi}^2 \phi^2 \ , \\
V_{II}&=&\frac{3m_{\phi}^2}{2} \big(1- e^{-\frac{2\phi}{\sqrt{3}}}\big)^2 \ .
\label{pots}
\end{eqnarray}

To see the relation with $R^2$ gravity, we recall
its action in the Jordan frame
\begin{equation}
S[g_{\mu\nu}^*] = 
\frac{1}{16\pi} \int d^4x  \sqrt{-g^*} \big( R^* +  {a} {R^*}^2 + L_{M} (g_{\mu\nu}^*, \chi) \big)
~,
\label{fR}
\end{equation} 
where $R^*$ is the Ricci scalar associated with the metric $g_{\mu\nu}^*$.
The parameter ${a}$ is a positive constant with units of
$[length]^{2}$, and controls the strength of the $R^2$ deviation 
from General Relativity. 
This theory can be recast into a particular 
Brans-Dicke STT with Jordan frame action
\citep{Yazadjiev:2014cza}, \citep{Yazadjiev:2015zia}
\begin{equation}
S[g_{\mu\nu}^*,\psi] = \frac{1}{16\pi }\int d^4 x \sqrt{-g^*} \big(\psi R^* - U(\psi) + L_{M} (g_{\mu\nu}^*, \chi) \big)
~,
\label{JordanAction}
\end{equation}
with scalar field $\psi$, and
potential $U(\psi)$
\begin{equation}
U(\psi) = R^* \psi - f(R^*) = \frac{1}{4 {a}} (\psi - 1)^2~,
\quad \psi = \frac{df}{dR^*}
~.
\end{equation}
The transition to the Einstein frame follows 
with help of the relations
\begin{equation}
g^*_{\mu\nu} = A^2 g_{\mu\nu}
\,\Rightarrow \, R^* = A^{-2} R - 6 A^{-3} \nabla_{\mu} \left(\partial^{\mu} A\right)
~, \quad A^{-2} = \psi = e^{\frac{2}{\sqrt{3}}\phi}
~.
\end{equation}
In the Einstein frame with the new metric $g_{\mu\nu}$ 
and the scalar field $\phi$ we then obtain
the action (\ref{EinsteinAction})
with the potential
\begin{equation}
V(\phi)= \frac{1}{4 {a}} \big(1- e^{-\frac{2\phi}{\sqrt{3}}}\big)^2
~.
\end{equation}
The potentials in the two frames are related by 
\begin{equation}
    U(\psi)= A^{-4} V(\phi) \ .
\end{equation}

This action (\ref{EinsteinAction}) now has an 
explicit kinetic term for the scalar field.
The parameter ${a}$ is related to the mass of the scalar,
\begin{equation}
    m_{\phi} = \frac{1}{\sqrt{6 {a}}} \ .
\end{equation}
Note that, when the scalar field is weak, 
the potential is essentially quadratic, given by $V(\phi)\sim\frac{\phi^2}{3 {a}}$.
When ${a}$ is set to zero, 
General Relativity is recovered with an 
infinitely massive scalar field,
that is, hence, suppressed to zero.
In the action in the Einstein frame (\ref{EinsteinAction})
the matter and scalar field are non-minimally coupled. 
In contrast, in the Jordan frame (\ref{JordanAction})
the scalar field is non-minimally coupled 
to the Ricci scalar,
which makes the action highly non-linear.
For further discussions on the Einstein and Jordan frames
in STTs, see e.g., \citep{Faraoni:1999hp,Bhattacharya:2017pqc}.

In this paper, we are working 
in the Einstein frame, and thus  
with the action (\ref{EinsteinAction}). 
The field equation for the metric $g_{\mu\nu}$ is then given by,
\begin{equation}
G_{\mu\nu} = T^{(S)}_{\mu\nu} + 8 \pi T^{(M)}_{\mu\nu}
-\frac{1}{2}V(\phi)g_{\mu\nu}
~,
\label{eq_G}
\end{equation}
with Einstein tensor $G_{\mu\nu} = R_{\mu\nu} - \frac{1}{2}Rg_{\mu\nu}$.
The energy-momentum tensor of the scalar field $T^{(S)}$ is given by 
\begin{equation}
T^{(S)}_{\mu\nu}=2\partial_{\mu}\phi\partial_{\nu}\phi -
g_{\mu\nu} \partial^{\sigma}\phi\partial_{\sigma}\phi
\ ,
\end{equation}
and the energy-momentum tensor of the matter $T^{(M)}$, 
which is assumed to be a perfect fluid, is
\begin{equation}
T^{(M)}_{\mu\nu} = (\rho + p)u_{\mu}u_{\nu} + pg_{\mu\nu}
~,
\label{T_matter}
\end{equation}
where $\rho$ is the energy density, $p$ is the pressure,
and $u_{\mu}$ is the 4-velocity of the fluid.
We assume the existence of a barotropic equation of state
relating the energy density and the pressure, 
determined by the properties of matter at high densities.
Since this relation is typically calculated 
in the equivalent (physical) Jordan frame, 
we need to transform the energy density and the pressure 
from the Jordan frame to the Einstein frame.
The relations with the pressure $\hat{p}$ and the density
$\hat{\rho}$ defined in the Jordan frame are
\begin{equation}
p = A^4 \hat{p}~, \quad \rho = A^4 \hat{\rho} ~,
\label{p_rho_EJ}
\end{equation}
and the equation of state is a relation of the form
$\hat{\rho}=\hat{\rho}(\hat{p})$.
The field equation for the scalar field in (\ref{EinsteinAction}) is
\begin{equation}
\nabla_{\mu}\nabla^{\mu}\phi = -4\pi\frac{1}{A}\frac{dA}{d\phi} T^{(M)} + \frac{1}{4} \frac{dV}{d\phi}
~,
\label{scalar_eom}
\end{equation}
where $T^{(M)}$ is the trace of the energy-momentum tensor of the matter.

\vspace{0.5cm}

\subsection{Static neutron stars with scalar hair}

For static and spherically symmetric neutron stars 
we consider the following Ansatz for the metric 
\begin{equation}
ds^2 = g_{\mu\nu}^{(0)} dx^{\mu} dx^{\nu} = -e^{2\nu(r)} dt^2
+ e^{2\lambda(r)} dr^2 + r^2 (d\theta^2 + 
\text{sin}^2 \theta \, d\varphi^2 )
~.
\end{equation}
The scalar field, energy density and pressure are simply
given by $\phi=\phi_0(r)$, $\hat{\rho}=\hat{\rho}_0(r)$ 
and $\hat{p}=\hat{p}_0(r)$, respectively, 
and the four-velocity of the static fluid 
is given by
$u^{(0)}=-e^{\nu}dt$.

Inside the star, the equations for the static functions are
\begin{eqnarray}
&&\frac{1}{r^2}\frac{d}{dr}\left[r(1- e^{-2\lambda})\right]= 8\pi
A_0^4 {\hat{\rho}_0} + e^{-2\lambda}\left(\frac{d\phi_0}{dr}\right)^2
+ \frac{1}{2} V_0,  \label{eq_lambda} \\
&&\frac{2}{r}e^{-2\lambda} \frac{d\nu}{dr} - \frac{1}{r^2}(1-
e^{-2\lambda})= 8\pi A_0^4 \hat{p}_0 +
e^{-2\lambda}\left(\frac{d\phi_0}{dr}\right)^2 - \frac{1}{2}
V_0, \label{eq_nu}
\\
&&\frac{d\hat{p}_0}{dr}= - (\hat{\rho}_0 + \hat{p}_0) \left(\frac{d\nu}{dr} +\frac{1}{A_0}\frac{dA_0}{d\phi_0}\frac{d\phi_0}{dr} \right), \label{eq_pJ} 
\\
&&\frac{d^2\phi_0}{dr^2} + \left(\frac{d\nu}{dr} -
\frac{d\lambda}{dr} + \frac{2}{r} \right)\frac{d\phi_0}{dr}= 4\pi\frac{1}{A_0}\frac{dA_0}{d\phi_0} A_0^4(\hat{\rho}_0-3\hat{p}_0)e^{2\lambda} + \frac{1}{4}
\frac{dV_0}{d\phi_0} e^{2\lambda}, \label{eq_phi} 
\end{eqnarray}
where $A_0=A(\phi_0)$ and $V_0=V(\phi_0)$.

The second order differential equation
(\ref{eq_phi}) for $\phi$ is obtained from
the scalar field equation
(\ref{scalar_eom}) at the static level,
while the others come from the Einstein
equations (\ref{eq_G}).
The system of equations has to be complemented with an
equation of state $\hat{\rho}_0=\hat{\rho}_0(\hat{p_0})$. 
Note that the static Einstein frame
quantities $p_0 = A(\phi_0)^4 \hat{p}_0$ 
and $\rho_0 = A(\phi_0)^4 \hat{\rho}_0$, 
will appear in some formulas below 
(to simplify notation).

\begin{figure}[t!]
	\centering
	\includegraphics[width=.38\textwidth, angle =-90]{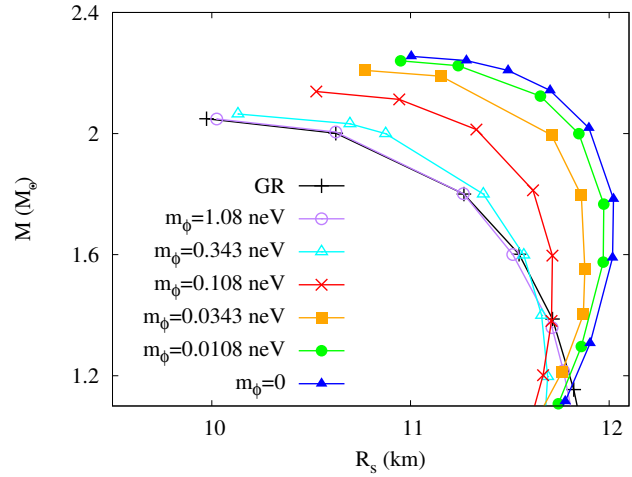}
	\caption{
	Mass $M$ in solar masses $M_\odot$
	versus physical radius $R_s$ in km. The
		colors represent different values of the scalar field
		mass $m_{\phi}$, with the general relativistic limit
		in black. 
	}
	\label{fig:background}
\end{figure}

Solutions describing neutron stars have to satisfy certain boundary conditions. At the center of the star, the configuration has to be regular. An expansion yields
\begin{eqnarray}
\lambda = \left[\frac{4}{3}\pi A_c^4 \hat{\rho}_c +\frac{1}{12} V_c  \right] r^2 + o(r^4), \\
\nu = \nu_c + \left[\frac{2}{3}\pi A_c^4 (3\hat{p}_c+\hat{\rho}_c) -\frac{1}{12} V_c  \right] r^2 + o(r^4), \\
\hat{p} = \hat{p}_c + (\hat{p}_c+\hat{\rho}_c)\left[
\frac{2\pi A_c^3}{3\sqrt{3}} \frac{dA_0}{d\phi_0}\biggr\rvert_c(3\hat{p}_c-\hat{\rho}_c)+\frac{\sqrt{3}}{72}\frac{dV_0}{d\phi_0}\biggr\rvert_c + \left(\frac{V_c}{12}-\frac{2\pi}{3}A_c^4(3\hat{p}_c+\hat{\rho}_c)\right)
 \right] r^2 + o(r^4), \\
 \phi_0 = \phi_c + \left[
 \frac{2\pi}{3}A_c^3 \frac{dA_0}{d\phi_0}\biggr\rvert_c(3\hat{p}_c-\hat{\rho}_c)+\frac{1}{24}\frac{dV_0}{d\phi_0}\biggr\rvert_c
 \right] r^2 + o(r^4), 
\end{eqnarray}
where $\nu_c=\nu(0)$, $\hat{p}_c=\hat{p}(0)$,
$\phi_c=\phi_0(0)$, $V_c=V_0(\phi_c)$, $A_c=A_0(\phi_c)$, 
etc. This expansion has only three free parameters $\nu_c$,
$\hat{p}_c$ and $\phi_c$, the others being determined by the
couplings and the equation of state. However, a global
solution that is asymptotically flat will have only one free
parameter at the center (typically chosen to be the central
pressure $\hat{p_c}$ in the numerical calculations).

The border of the neutron star is defined by the point 
$r=r_s$, where the pressure vanishes, $\hat{p}_0(r_s)=0$. The physical (Jordan frame) radius of the star is given by $R_s=A(\phi_0(r_s))r_s$, see \citep{Blazquez-Salcedo:2018qyy} for more details.
Outside the neutron star there is no fluid, thus
$\hat{p}_0=\hat{\rho}_0=0$ in equations 
(\ref{eq_lambda}-\ref{eq_phi}). 
We are interested in asymptotically flat solutions 
with vanishing scalar field at infinity.
However, for ${a}\neq 0$, the scalar field outside the star
does not vanish. Nonetheless, sufficiently far from the star,
the scalar field decays exponentially with 
$\phi \sim \frac{1}{r}e^{-m_{\phi} r}$, 
and the background metric is essentially given by the
Schwarzschild solution, with 
$e^{2\nu}=e^{-2\lambda}\sim 1-2M/r$, 
where $M$ is the total mass of the neutron star.

The mass--radius relation for most of the 
background neutron star configurations
investigated in the following is exhibited in Figure 
\ref{fig:background}, where
the mass $M$ is given in solar masses $M_\odot$
and the physical (Jordan frame) radius $R_s$ in km. 
The chosen potential is $V_{II}$ ($R^2$ gravity),
and the scalar field mass $m_\phi$ assumes several 
values in the physically interesting mass range
\citep{Naf:2010zy,Brito:2017zvb}.
The color coding is 
purple, cyan, red, orange and green
for $m_{\phi}=1.08$, $0.343$, $0.108$, $0.0343$ 
and $0.0108$ neV, respectively. 
Also shown are the general relativistic limit (black),
and the case of a massless scalar field (blue).
For the SLy equation of state the maximum mass of
static neutron stars in General Relativity 
is slightly above two solar masses.
In the STTs studied, the value of the maximum mass 
is increased the more, the smaller the scalar field mass is,
reaching about 2.25 solar masses.
At the same time, the neutron star radii become larger 
except for rather small neutron star masses.

\section{Perturbations in massive scalar-tensor theory} 
\label{sec_QNM}

\vspace{0.5cm}

\subsection{Setup}

In this section we present the quasinormal mode formalism, focusing on polar perturbations. 
In the Einstein frame, we perturb the background metric, the scalar field and the fluid in the following way:
\begin{eqnarray}
g_{\mu\nu} = g_{\mu\nu}^{(0)}(r) + \epsilon h_{\mu\nu}(t,r,\theta,\varphi)~, \\
\label{Phiperturb}
\phi = \phi_{0}(r) + \epsilon \delta\phi (t,r,\theta,\varphi)~, \\
\rho = \rho_0 (r) + \epsilon \delta\rho(t,r,\theta,\varphi)~,  \\
p = p_0(r) + \epsilon \delta p(t,r,\theta,\varphi)~, \\
u_{\mu} = u^{(0)}_{\mu} (r) + \epsilon \delta u_{\mu} (t,r,\theta,\varphi)
~,
\end{eqnarray}
where $\epsilon<<1$ is the perturbation parameter. 
The zeroth order of the configurations describes a static and spherically symmetric background, discussed in the previous section, while the perturbations are in general dependent on time, radial coordinate and angular directions. 
 
Assuming that the perturbation functions 
can be expanded as a product of radial, temporal and angular components, 
they can be further separated into classes of axial and polar perturbations,
depending on the transformation of the angular component under parity
\citep{ReggeW,Zerilli:1970se,1967ApJ...149..591T,1969ApJ...155..163P,1969ApJ...158..997T,1970ApJ...159..847C,Thorne:1980ru,Detweiler:1985zz,doi:10.1098/rspa.1991.0016,doi:10.1098/rspa.1991.0117,doi:10.1098/rspa.1991.0104,PhysRevD.43.1768,Kojima:1992ie,FernandezJambrina:2003mv}.
For polar perturbations, the spherical harmonics transform as
$
Y_{lm}(\theta,\varphi) \rightarrow Y_{lm}(\pi-\theta,\pi+\varphi) =
(-1)^{l}Y_{lm}(\theta,\varphi)$.
For a study of the current theory in the axial case we refer the reader to \citep{Blazquez-Salcedo:2018qyy}.

The Ansatz for the polar perturbations of the metric is
\begin{equation}
h_{\mu\nu}^{(\text{polar})} = \sum\limits_{l,m}\,\int    
\left[
\begin{array}{c c c c}
r^l e^{2\nu} H_0 Y_{lm} & -i \omega r^{l+1} H_1 Y_{lm} & 
0 & 0 \\
-i \omega r^{l+1} H_1 Y_{lm} &  r^l e^{2\lambda} H_2 Y_{lm} & 
0 & 0 \\
0 & 0
& r^{l+2} K  Y_{lm} & 0
\\
0 & 0  & 
0 & r^{l+2}  \sin^2\theta K  Y_{lm} \\
\end{array}
\right]
e^{-i\omega t} d\omega
~,
\end{equation}
in the order of $(t,r,\theta,\varphi)$ 
in the rows and columns of the matrix. 
The functions $H_0, H_1, H_2, K$ 
only depend on the radial coordinate $r$, 
the integer multipole numbers $l$, $m$, 
and the complex wave frequency $\omega$, where
$\omega = \omega_{R} + i\omega_{I}$ for $\omega_R, 
\omega_I \in \mathbb{R}$.
The functions $Y_{lm}$ are the standard spherical harmonics.
The scalar field can be decomposed as
\begin{equation} 
\delta \phi =  \sum\limits_{l,m}\,\int  r^l \phi_1 \, 
Y_{lm} e^{-i\omega t} d\omega ~,
\end{equation}
where again the function $\phi_1$ 
depends on $r,l,m$ and $\omega$.

Concerning the perturbation of the fluid inside the star,
the scalar quantities, such as the energy density $\rho$ 
and pressure $p$, can be decomposed similarly 
to the scalar field
\begin{equation}
\delta \rho = \sum\limits_{l,m}\,
\int   r^l E_{1 } Y_{lm} e^{-i\omega t }d\omega~, \quad
\delta p = \sum\limits_{l,m}\,
\int   r^l \Pi_{1}  Y_{lm} e^{-i\omega t} d\omega~, 
\end{equation}
while the perturbation of the 4-velocity is given by 
\begin{eqnarray}
\delta u_{\mu} = 
\sum\limits_{l,m}\,\int    
\left[
\begin{array}{c}
\frac{1}{2} r^l e^{\nu} H_0 Y_{lm}  \\
r^l i\omega e^{-\nu} 
\left(e^{\lambda}W/r -r H_1 \right) Y_{lm}  \\
-i\omega r^l e^{-\nu} V \partial_{\theta} Y_{lm}
\\
-i\omega r^l e^{-\nu} V \partial_{\phi} Y_{lm}  \\
\end{array}
\right]
e^{-i\omega t} d\omega
~,
\end{eqnarray}
The functions $V$ and $W$ depend on $r,l,m,\omega$.

Before continuing, we note that although we have given 
the perturbations in the Einstein frame, 
they can be alternatively defined in the Jordan frame. 
Perturbations in the Jordan frame for the scalar 
and the metric 
($\delta\psi$ and $h_{\mu\nu}^*$, respectively) 
can be expressed as a combination of the perturbations 
in the Einstein frame,
\begin{equation}
\delta\psi = \frac{2}{\sqrt{3}}
e^{\frac{2}{\sqrt{3}}\phi_0}\delta\phi~,
\quad
h_{\mu\nu}^* = e^{-\frac{2}{\sqrt{3}}\phi_0} 
( h_{\mu\nu} - \frac{2}{\sqrt{3}} 
g_{\mu\nu}^{(0)}\delta\phi)
~.
\end{equation}
Concerning the energy density and pressure, 
the barotropic equation of state in the Jordan frame 
implies a relation between the perturbations 
$\delta p$, $\delta \rho$ and $\delta \phi$
\begin{eqnarray}
\label{baro_jordan}
\delta\rho = \frac{d\hat{\rho}_0}{d\hat{p}_0}\delta p 
+ 4 A_0^3 \frac{dA_0}{d\phi_0}(\hat{\rho}_0-\hat{p}_0
\frac{d\hat{\rho}_0}{d\hat{p}_0})\delta\phi 
~.
\end{eqnarray}

\vspace{0.5cm}

\subsection{Equations of the polar perturbations}

Employing the Ansatz shown in the previous section 
for the perturbations on the field equations 
(\ref{eq_G}) and (\ref{scalar_eom}), 
leads to a system of  
ordinary differential equations in $r$, 
that is characterized by the eigenvalue $\omega$, 
and the multipole number $l$, 
but that is independent of $m$ 
because of the spherical symmetry. 

The modified Einstein equation (\ref{eq_G}) 
result in six ordinary differential equations
\begin{eqnarray}
\label{ed_d2K}
{\frac {{d}^{2}{K}}{{d}{r}^{2}}}  
=-8\pi {{e}^{2\lambda  }}{E_1}
+\frac{1}{2r^2}{{ \left( 2 
\left( {\frac {d\phi_0}{{d}r}} \right)^{2}{r}^{2}
-4r{\frac {d\lambda}{{d}r}}+({l}{{e}^{2\lambda}}+2)(l+1)
\right) {H_2}  }}
 \nonumber \\
+2{\frac {d\phi_0}{{d}r}}{\frac {d\phi_1}{{d}r}} 
+\frac{1}{6r}{{ \left( 
		\sqrt{3}e^{2\lambda}\frac{r}{a}
		(
		{{e}^{-\frac{4}{\sqrt{3}}\phi_0  }}
		-{{e}^{-\frac{2}{\sqrt{3}}\phi_0 }}
		)
		-12l {\frac {d\phi_0}{{d}r}}\right) \phi_1  }}  
\nonumber \\
+\frac{1}{2r^2}{{\left( 2  lr  {\frac
{d\lambda}{{d}r}}+{{e}^{2\lambda  }}(l(l+1)-2)-2l(l+2) 
\right) {K}  }}
+{\frac { \left(   r{\frac {d\lambda}{{d}r}}    -2l-3 \right)
}{r}}{\frac {d{K}}{{d}r}}+{\frac {1}{r}}
{\frac {d{H_2}}{{d}r}} 
  \nonumber \\
+\frac{1}{8r^2}{{ \left( 64\pi \rho  {r}^{2}{{e}^{2\lambda}}
+8 \left( {\frac {d\phi_0}{{d}r}}   \right) ^{2}{r}^{2}
		+e^{2\lambda}\frac{{r}^{2}}{a}
		({{e}^{-\frac{2}{\sqrt{3}}\phi_0}}-1)^2
		-16r {\frac {d\lambda}{{d}r}}
		+8(1-e^{2\lambda})\right) {H_0}}} 
~,
\end{eqnarray}
\begin{eqnarray}
\label{eq_dK}
{\frac {d{K}}{{d}r}}  ={\frac {1 }{r}}{H_2}+{\frac { \left({\frac {d\nu}{{d}r}} r-l-1\right) }{r}}{K}-2{\frac {d\phi_0}{{d}r}}\phi_1  -8\pi{\frac {{{e}^{\lambda}} \left(p+\rho\right)}{r}}{W}  
\nonumber \\
+\frac{1}{8r}{{ \left(
		4l(l+1)-16{\frac {d\lambda }{{d}r}}  r{{e}^{-2\lambda }}+8({{e}^{-2\lambda}}-1)\right) {H_1}  }}
\nonumber \\
+\frac{1}{8r}{{ \left( 64\pi \rho {r}^{2}
		-\frac{r^2}{a}
		({{e}^{-\frac{2}{\sqrt{3}}\phi_0}}-1)^2
		+8 \left( {\frac {d\phi_0}{{d}r}}   \right) ^{2}{r}^{2}{{e}^{-2\lambda  }}
		\right) {H_1}  }} 
		~,
\end{eqnarray}
\begin{eqnarray}
\label{eq_dH1}
{\frac {d{H_1}}{{d}r}}  ={\frac {  {{e}^{2\lambda  }
	}}{r}}{H_2}+{ \left(  {\frac {d\lambda}{{d}r}}
		   -  {\frac {d\nu}{{d}r}}
		  -\frac{l+1}{r} \right)  }{H_1}+{\frac {  {{e}^{2
				\lambda  }}}{r}}{K}-\frac{16\pi}{r}{{   {{e}^{2\lambda  }} \left( p
		 +\rho   \right) }}{V}
		 ~,
\end{eqnarray}
\begin{eqnarray}
\label{eq_dH0}
{\frac {d{H_0}}{{d}r}}  =-8\pi {{e}^{2\lambda  }}r{\Pi_1} -2r{\omega}^{2}{{e}^{-2\nu  }}{H_1}  +\frac{l}{2r}{{\left( (l+1){{e}^{2\lambda  }}-2 \right) {H_0}  }} -2r {\frac {d\phi_0}{{d}r}}{\frac {d\phi_1}{{d}r}}  \nonumber \\ -\frac{1}{8r}{{ \left( 64\pi {{e}^{2\lambda  }}p  {r}^{2}-\frac{r^2}{a}e^{2\lambda}
		({{e}^{-\frac{2}{\sqrt {3}}\phi_0}}-1)^2
		+8{{e}^{2\lambda  }} \right) {H_2} }} \nonumber \\
+\frac{1}{2r}{{ \left( 2{{e}^{2(\lambda-\nu)  }}{\omega}^{2}{r}^{2}-{{e}^{2\lambda  }}(l-1)(l+2)+2l \left(r{\frac {d\nu}{{d}r}}+1\right) \right) {K} }} \nonumber \\
 + \left( 
 \frac{1}{2\sqrt{3}}\frac{r}{a} e^{2\lambda} (
 {{e}^{-\frac{2}{\sqrt {3}}\phi_0}}-{{e}^{-\frac{4}{\sqrt{3}}\phi_0}})-2l {\frac {d\phi_0}{{d}r}} \right) \phi_1  + \left( r{\frac {d\nu}{{d}r}}+1 \right){\frac {d{K}}{dr} }
 ~,
\end{eqnarray}
\begin{eqnarray}
\label{eq_dH0_2}
{\frac {dH_0}{{d}r}}=-{{
		\left(  {\frac{d\nu}{{d}r}}  
	 +\frac{1}{r} \right) {H_2}}}-r{\omega}^{
	2}{{e}^{-2\nu  }}{H_1} -{ { \left(  {\frac {d\nu}{{d}r}}+\frac{l-1}{r} \right) {H_0}  }}
+{\frac {l}{r}}{K}+4 {\frac {d\phi_0}
	{{d}r}} \phi_1  +{
	\frac {d{K}}{{d}r}}  
	~,
\end{eqnarray}
\begin{eqnarray}
\label{eq_d2K_2}
{\frac {{ d}^{2}{K }}{{ d}{r}^{2}}} 
-{\frac {{ d}^{2}{H_0 }}{{ d}{r}^{2}}} =16\pi {{ e}^{2\lambda  }}{\Pi_1 }  -4  \left( {\frac {d\phi_0}{{ d}r}}   \right) {\frac { d\phi_1}{{ d}r}} + \left( {\frac { d\nu}{{ d}r}}  +\frac{1}{r} \right) {\frac { d{H_2 }}{{ d}r}}
+2 r{\omega}^{2}{{ e}^{-2 \nu  }}{\frac { d{H_1 }}{{ d}r}} 
 \nonumber  \\
-2 {{ e}^{-2 \nu  }}{\omega}^{2} \left( r{\frac { d\lambda}{{ d}r}}-l-2 \right) {H_1 }  -\frac{1}{2{{r}^{2}}}  {l \left( 2r  {\frac { d\lambda}{{ d}r}}-4r   {\frac { d\nu}{{ d}r}}+l{{ e}^{2 \lambda  }}+{{ e}^{2 \lambda  }}-2 l \right) {H_0 }  }
 \nonumber  \\
 +\frac{1}{2{{r}^{2}}} { \left( -2{{ e}^{-2\nu  +2\lambda  }}{\omega}^{2}{r}^{2}+4 \left( {\frac { d\phi_0}{{ d}r}}   \right) ^{2}{r}^{2}-4{r}^{2} {\frac { d\lambda}{{ d}r}} {\frac { d\nu}{{ d}r}}  +4 \left( {\frac { d\nu}{{ d}r}}   \right) ^{2}{r}^{2}+4{r}^{2}{\frac {{ d}^{2}\nu}{{ d}{r}^{2}}}  \right) {H_2 }  }
 \nonumber  \\
  +\frac{1}{2{{r}^{2}}} { \left(   +{l}^{2}{{ e}^{2\lambda  }}+2lr {\frac { d\nu}{{ d}r}}   -4 r {\frac { d\lambda}{{ d}r}}+l{{ e}^{2 \lambda }}+4r {\frac { d\nu}{{ d}r}} +2 l \right) {H_2 }  }
  \nonumber \\
 +\frac{1}{4r^{2}} { \left( 64 \pi  {{ e}^{2 \lambda  }}p  {r}^{2}-4 {{ e}^{-2 \nu  +2 \lambda  }}{\omega}^{2}{r}^{2}+8 {r}^{2}{\frac { d\lambda}{{ d}r}}{\frac { d\nu}{{ d}r}}  -8  \left( {\frac { d\phi_0}{{ d}r}}   \right) ^{2}{r}^{2}-8  \left( {\frac { d\nu}{{ d}r}}   \right) ^{2}{r}^{2} \right) {K } } 
  \nonumber \\
  +\frac{1}{4r^{2}} { \left(
	4r(l+2) \left( {\frac { d\lambda}{{ d}r}} - {\frac { d\nu}{{ d}r}}   \right)
	-8 {r}^{2}{\frac {{ d}^{2}\nu}{{ d}{r}^{2}}} -4l(l+1)-\frac{{r}^{2}}{a} e^{2\lambda}
	(1-{{ e}^{-\frac{2}{\sqrt {3}}\phi_0  }})^2
	 \right) {K } } 
 \nonumber \\
  +\frac{1}{3{r}} {{ \left( 
  		\sqrt{3}\frac{r}{a} e^{2\lambda}\left(
  		{{ e}^{-\frac{4}{\sqrt {3}}\phi_0  }}-{{ e}^{-\frac{2}{\sqrt {3}}\phi_0  }}\right)-12  l {\frac { d\phi_0}{{ d}r}} \right) \phi_1  }}
 \nonumber \\
   + \left( \frac{1}{r}(2l+1)-{\frac { d\lambda}{{ d}r}}  +2 {\frac { d\nu}{{ d}r}}   \right) {\frac { d{H_0 }}{{ d}r}}  + \left( {\frac { d\lambda}{{ d}r}}  -{\frac { d\nu}{{ d}r}}  -{\frac {2}{r}}(l+1) \right) {\frac { d{K }}{{ d}r}}  
   ~.
\end{eqnarray}
In addition, it imposes the constraint
\begin{eqnarray}
\label{eq_H2}
H_2 = H_0
~.
\end{eqnarray}
The scalar field equation (\ref{scalar_eom}) results in
\begin{eqnarray}
\label{eq_d2phi1}
{\frac {{ d}^{2}\phi_1}{{ d}{r}^{2}}} =
-\frac{4}{\sqrt {3}}\pi{{ e}^{2\lambda  }}{(E_1-3\Pi_1)}
+{\frac{l}{2r}}{\frac {d\phi_0}{{d}r}}{H_0}
+{\frac { d\phi_0}{{ d}r}}{{ e}^{-2\nu}}{\omega}^{2}r{H_1}
-{\frac {l}{r}}{\frac {d\phi_0 }{{ d}r}} K
\nonumber \\
+\frac{1}{6r^2}{{\left(
		-e^{2\lambda}r^2\left(
	 6{{e}^{-2\nu  }}{\omega}^{2}+({{e}^{-\frac{2}{\sqrt{3}}\phi_0}}-2{{e}^{-\frac{4}{\sqrt{3}}\phi_0 }})\frac{1}{a}
		\right)
		+6lr\left({\frac{d\lambda}{{d}r}} -{\frac{d\nu}{{d}r}} \right)+6l(1+l)({{e}^{2\lambda}}-1)\right)\phi_1}}
\nonumber \\
+\frac{1}{2r}{{ \left( -2r {\frac {d\phi_0}{{ d}r}}   {\frac { d\lambda}{{ d}r}}  +2r {\frac { d\phi_0}{{d}r}} {\frac { d\nu}{{ d}r}} +2r{\frac {{ d}^{2}\phi_0}{{ d}{r}^{2}}} + {\frac { d\phi_0}{{ d}r}} l+4{\frac { d\phi_0}{{ d}r}}  \right) {H_2}  }}
\nonumber \\
+{{\left({\frac {d\lambda}{{d}r}}  -{\frac{d\nu}{{d}r}}-\frac{2}{r}(l+1)\right)  }} {\frac {d\phi_1}{{ d}r}}+\frac{1}{2}{\frac{d\phi_0}{{ d}r}}\left({\frac {d{H_2}}{{ d}r}}  +{	\frac {d{H_0}}{{ d}r}}  -2{\frac 	{ d{K}}{{ d}r}}   \right) 
~.
\end{eqnarray}
The barotropic condition on the equation of
state (\ref{baro_jordan}) becomes a
relation between the energy density
perturbation, pressure perturbation and
scalar field perturbation,
\begin{eqnarray}
\label{eq_E1}
{E_1}= \frac{d\hat{\rho}_0}{d\hat{p}_0} {\Pi_1}
 -\frac{4}{\sqrt{3}}{{\rm e}^{-\frac{4}{\sqrt{3}}\phi_0
		}} \left( \hat{\rho}_0-\hat{p}_0 \frac{d\hat{\rho}_0}{d\hat{p}_0}\right) \phi_1 
		~.
\end{eqnarray}

By tedious algebraic manipulations
the nine equations (\ref{ed_d2K})-(\ref{eq_E1})
can be simplified.
For this purpose
it is convenient to define the following function 
of the perturbations \citep{Lindblom:1983ps},
\begin{eqnarray}
X={\omega}^{2} \left(\hat{p}_0 +\hat{\rho}_0  
\right) {{e}^{-\nu}}{V} 
-\frac{1}{r}{\frac{d\hat{p}_0}{{d}r}} {{e}^{\nu-\lambda}}{W}+\frac{1}{2}\left( \hat{p}_0+\hat{\rho}_0\right) {{e}^{\nu}}{H_0} 
~.
\end{eqnarray}
The resulting minimal system of differential equations 
is given by a set of six first order differential equations
for the functions 
$\Psi = (K,  H_1, W, X, \phi_1, \frac{d\phi_1}{dr})$, 
which takes the form
\begin{equation}
\label{eq_Psi}
\frac{d}{dr} \Psi + \sigma\Psi = 0~,
\end{equation}
where $\sigma$ is a matrix that depends in a complicated way
on the static functions 
$\nu, \lambda, \phi_0, \hat{p}_0, \hat{\rho}_0$, 
and also on the eigenvalue $\omega$ 
and the multipole number $l$. 

Note that inside the star, the perturbation is described by
the functions $(K, H_1)$, which parametrize the metric
perturbation, $(W, X)$ which parametrize the fluid
perturbation and $(\phi_1,\frac{d\phi_1}{dr})$
which is the scalar field perturbation. 
Outside the star, the system simplifies, 
since $\hat{p}=\hat{\rho}=0$. 
For instance, $W=X=0$ when there is no fluid, 
and the system reduces to a system of four first order
differential equations for $(K,  H_1)$ (metric) 
and $(\phi_1, \frac{d\phi_1}{dr})$ (scalar field).

Let us note that in STTs all perturbation equations 
are coupled with each other, whereas in the 
general relativistic limit, the system decouples
	on one hand into the metric and fluid perturbations, 
	and on the other hand into the scalar field perturbation,
	which is simply governed by the minimally coupled scalar
	test field equation in the general relativistic limit.
	
\vspace{0.5cm}

\subsection{Boundary conditions and asymptotic behaviour}

At the center of the star we impose regularity of the perturbations. This means that at the center of the star the perturbation functions satisfy the constraints
\begin{eqnarray}
\label{center_expan}
l(l+1)H_1(0)-2lK(0)-16\pi A_c^4(\hat{p}_c+\hat{\rho}_c)W(0)=0,
 \nonumber  \\
l\Pi_1(0) + \frac{l}{\sqrt{3}}A_c^4(3\hat{p}_c-\hat{\rho}_c)\phi_1(0) + A_c^4 (\hat{p}_c+\hat{\rho}_c) \left(\frac{l}{2} K(0)- \omega^2 e^{-2\nu_c}  W(0)\right)=0,
\nonumber \\
\frac{le^{-\nu_c}}{\hat{p}_c+\hat{\rho}_c}X(0)-\frac{l}{2}K(0)=\left(
\frac{8\pi l}{9} A_c^4 (3\hat{p}_c+2\hat{\rho}_c)-e^{-2\nu_c}\omega^2-\frac{l}{72a}(e^{-\frac{4}{\sqrt{3}}\phi_c}-4e^{-\frac{2}{\sqrt{3}}\phi_c}+3)
\right)W(0)
,
 \nonumber \\
\frac{d\phi_1}{dr}(0)=0 \, , \, \, \,  lV(0)+W(0)=0 \, , \, \, \, H_2(0)-K(0)=0 \, , \, \, \,  H_0(0)-K(0)=0
.
\end{eqnarray}
At infinity the massive scalar field is exponentially
suppressed. Therefore the scalar
perturbation is effectively asymptotically decoupled 
from the metric perturbations. 
Consequently, sufficiently far from the star, 
the oscillation can be described 
by the standard Zerilli function $Z_g(r)$ given by
\begin{eqnarray}
H_1(r) = \frac{3M(nr+M)-n r^2}{r^{l+1}(2M-r)(nr+3M)}Z_g(r) + \frac{r}{r^l(r-2M)}F_g(r), \\
K(r) = \frac{nr^2(n+1)+3M(nr+2M)}{r^{l+2}(nr+3M)}Z_g(r) + \frac{1}{r^l}F_g(r),
\end{eqnarray}
where $n=l(l+1)/2$, and $F_g(r)$ is some 
supplementary function. 
Then the two first order equations for $(K, H_1)$ 
can be rewritten as a more standard Schr\"odinger-like
equation for $Z_g$
\begin{eqnarray}
\frac{d^2Z_g}{dy^2}= (G_g(r)-\omega^2) Z_g
\label{eq_d2Zg}
\end{eqnarray}
with tortoise coordinate $\frac{dy}{dr}=e^{\lambda-\nu}$ 
and the effective potential $G_g(r)$ 
for space-time perturbations
\begin{eqnarray}
G_g(r)=
\frac{2(r-2M)}{r^4(nr+3M)^2}\left(n^2(n+1)r^3+3Mn^2r^2+9M^2nr+9M^3\right)
~.
\end{eqnarray}
Note that asymptotically, the potential goes to zero, $G_g(r\to\infty)\to 0$. Therefore the asymptotic behaviour 
of the perturbation is given by the combination of an 
ingoing and an outgoing solution of the form
\begin{eqnarray}
Z_g \sim A_{in} e^{-i\omega y} + A_{out} e^{i\omega y} \ .
\end{eqnarray}

We now consider the scalar field. Since this component
decouples exponentially from the metric perturbation sufficiently far from the star, 
the scalar field equation can also be reformulated
as a Schr\"odinger-like equation.
Defining
\begin{eqnarray}
\phi_1(r)=r^l Z_s(r)
~,
\end{eqnarray}
the equation then becomes
\begin{eqnarray}
\frac{d^2Z_s}{dy^2}= (G_s(r)-\omega^2) Z_s \ 
\label{eq_d2Zs}
\end{eqnarray}
with the (same) tortoise coordinate $y$, 
and $G_s(r)$ denotes the effective potential 
for scalar perturbations
\begin{eqnarray}
G_s(r)=
\left(1-\frac{2M}{r}\right)
\left(
\frac{1}{r^2}
\left(
l(l+1)+\frac{2M}{r}
\right)
+\frac{1}{6a}
\right)
~.
\end{eqnarray}
Note that in this case, the potential goes asymptotically to $G_s(r\to\infty)\to \frac{1}{6a}=m_{\phi}^2$.
Therefore the scalar perturbation is asymptotically 
also given by a combination of outgoing and ingoing waves, 
but now of the form
\begin{eqnarray}
Z_s \sim A_{in} e^{-i\Omega y} + A_{out} e^{i\Omega y} \ ,
\end{eqnarray}
where $\Omega$ satisfies the following dispersion relation 
for the scalar field perturbations
\begin{eqnarray}
\Omega^2 = \omega^2 - m_{\phi}^2 \ .
\end{eqnarray}
The appearance of $\Omega$ is related to the fact 
that the scalar field
perturbations cannot propagate at the speed of light,
like the space-time perturbations, 
since for finite values of $a$, the scalar field is massive.
Only in the limit of $a\to\infty$ the scalar field 
becomes massless and we recover $\Omega=\omega$.

\vspace{0.5cm}

\subsection{Outline of the numerical method}

We now briefly summarize the numerical implementation 
of the quasinormal mode calculations. 
The very first step is the calculation 
of the background configurations. 
To this end we solve the static equations using Colsys
\citep{Ascher:1979iha}. The solutions are obtained 
by employing a compactified coordinate $x=\frac{r}{r_s+r}$, 
which allows to impose the physical boundary conditions
exactly at the center of the star, 
at its surface $r_s$ and at infinity. 
The input parameters are the central pressure $\hat{p}_c$
and the %
scalar field mass $m_\phi$.
The system has to be complemented with an equation of state. 
In this paper we shall focus on 
one particular choice for the equation of state, 
the SLy EOS \citep{Douchin:2001sv}, 
which is a representative model that captures
the basic features of realistic neutron star models.
We implement this equation of state by
using a piece-wise polytropic
approximation \citep{Read:2008iy}.

Once a particular background star with mass $M$ 
is calculated, it is used to calculate the coefficients 
of the matrix $\sigma$ of equation (\ref{eq_Psi}). 
Then we proceed to calculate the 
quasinormal modes for a particular configuration 
and a particular multipole number $l$. 
To do so, we solve the perturbation equations for 
a fixed value of $\omega$ in three different steps. 

First we obtain two independent solutions of 
equation (\ref{eq_Psi}) inside the star, 
satisfying the conditions (\ref{center_expan}), 
and imposing $X(r_s)=0$.
Second, these two solutions are continued outside the star, 
by requiring continuity of the metric 
and scalar field perturbations. 
These solutions are obtained up to some point $r_i>r_s$. 
We choose this $r_i$ so that $\phi_0(r_i)\lesssim 10^{-4}$.
In the third step, we calculate the phases of the metric 
and scalar perturbations in the background of a 
Schwarzschild solution with mass $M$. 
We impose purely outgoing wave solutions at infinity 
for both phases. To do so, we make use of the 
exterior complex scaling method on equations (\ref{eq_d2Zg}) 
and (\ref{eq_d2Zs}). For more details 
we refer the reader to
\citep{BlazquezSalcedo:2012pd,Blazquez-Salcedo:2013jka,Blazquez-Salcedo:2015ets,Blazquez-Salcedo:2018tyn,AltahaMotahar:2018djk,AltahaMotahar:2019ekm,Blazquez-Salcedo:2018qyy,Blazquez-Salcedo:2018pxo}.

Then we check if the phases obtained for the asymptotic
behaviour of the perturbation, which satisfy 
the outgoing wave behaviour, match with the full 
perturbative solution obtained in a region $0<r<r_i$. 
If they match, 
then $\omega$ is the eigenvalue of a quasinormal mode. 
If not, we repeat the process for different values 
of $\omega$ until matching is achieved.
In this way, we investigate the spectrum for numerous
configurations for several values of 
the scalar field mass $m_\phi$.

\section{Results} \label{sec_results}

Because of the nature of the polar perturbations, 
the spectrum of polar perturbations is much richer 
than the spectrum of axial perturbations, 
which involve only spacetime perturbations. 
Polar modes on the contrary feature 
different families of modes. 
In General Relativity
static and spherically symmetric neutron stars 
not only possess spacetime modes (w-modes), 
they also possess modes related to the fluid perturbation. 
For $l\geq 2$, the spectrum is dominated 
by the fundamental mode (f-mode), 
a nodeless fluctuation driven by pressure oscillations 
inside the star, which typically possesses 
the lowest frequency. 
But there are also excited pressure modes (p-modes) 
with higher values of the frequency. 

When considering radial perturbations in General Relativity,
stars are seen to possess 
a family of normal modes with $\omega^2 \in \mathbb{R}$,
that become unstable beyond the maximum mass of the
equation of state.
These modes are well known in the literature \citep{Chandrasekhar:1964zza,Chandrasekhar:1964zza,1966ApJ...145..505B,1966ApJ...145..514M,1977ApJ...217..799C,1983ApJS...53...93G,1992A&A...260..250V,1998IJMPD...7...49D}.
However, in General Relativity radial perturbations 
cannot propagate gravitational radiation 
outside the neutron star.

In STTs there is another degree of freedom, 
the scalar field $\phi$. Therefore
similarly to the scalar modes of hairy black holes 
\citep{Blazquez-Salcedo:2016enn,Blazquez-Salcedo:2017txk,Blazquez-Salcedo:2018pxo},
there are additional modes for scalarized neutron stars. 
The $l=0$ modes can propagate outside the neutron stars
in STTs, which makes them relevant for the study of
gravitational waves. 
In addition, of course, dipole ($l=1$) modes arise.
These and the $l=0$ modes are supported 
by the scalar field, but they are coupled via the field
equations with oscillations of the metric 
and the neutron star fluid.

\vspace{0.5cm}

\subsection{Quadrupole modes}

\begin{figure}
	\centering
	\includegraphics[width=.38\textwidth, angle =-90]{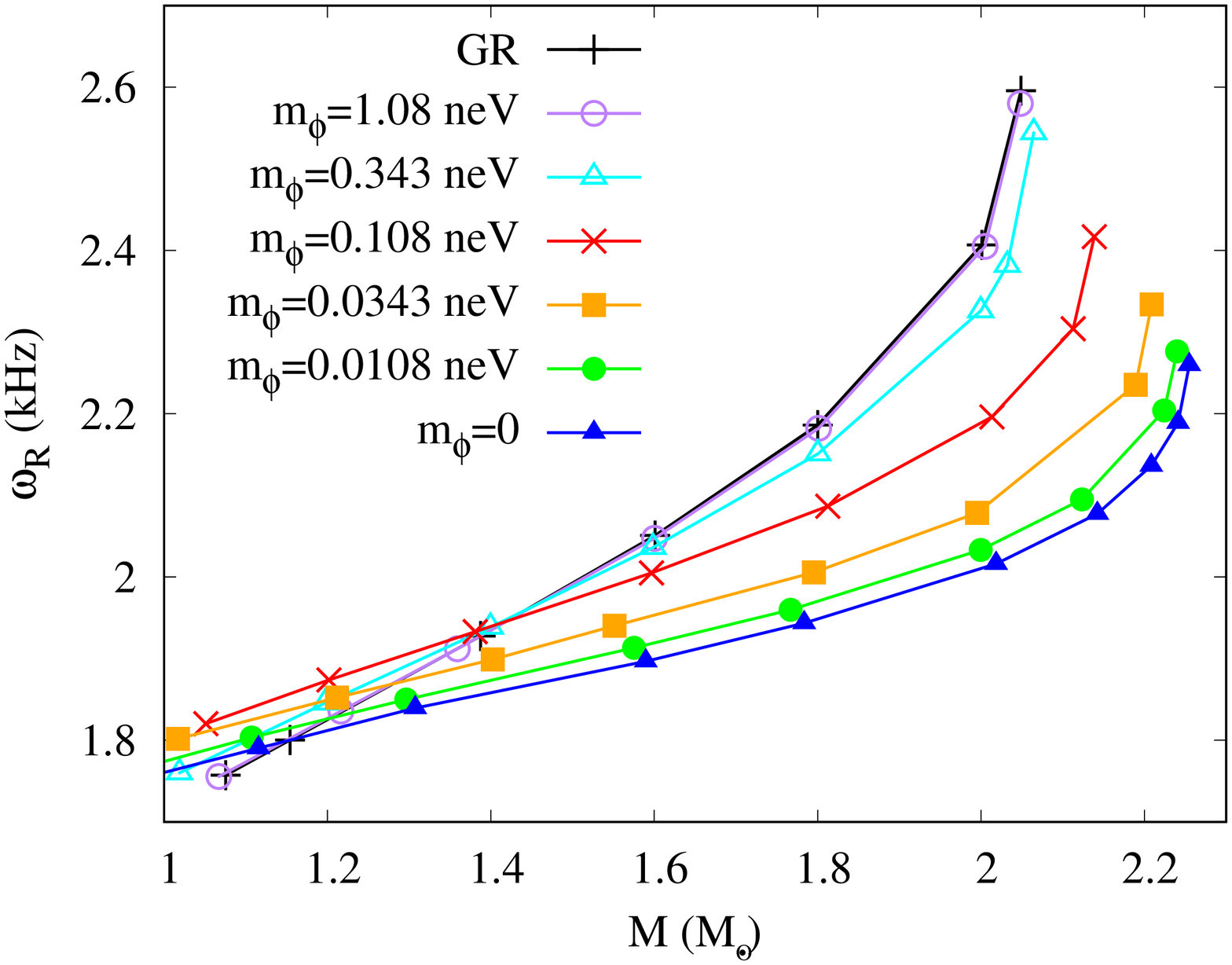}
	\includegraphics[width=.38\textwidth, angle =-90]{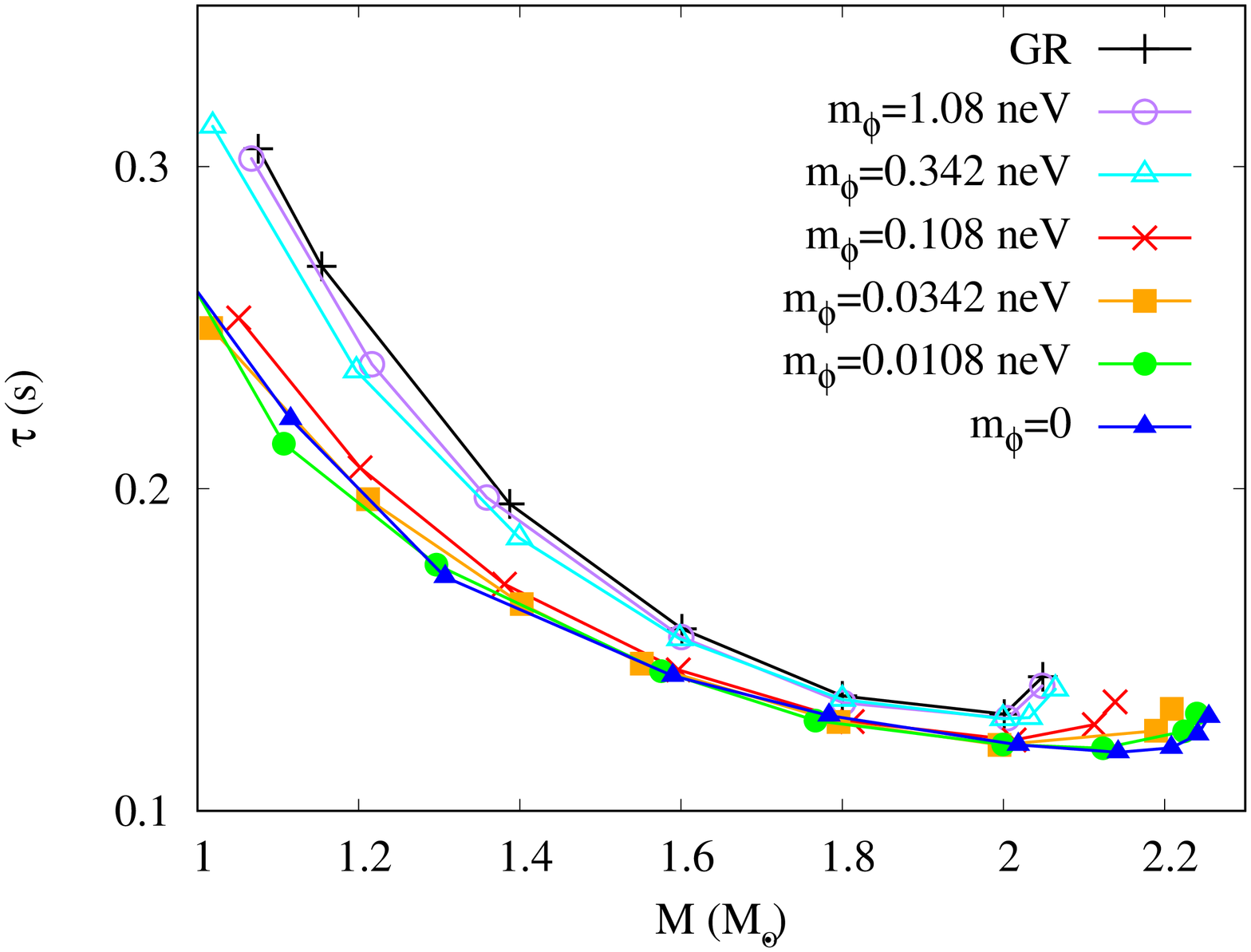}
	\caption{
		Frequency $\omega_R$ in kHz (left) and damping time $\tau$
		in s (right) versus the total
		mass of the neutron star $M$ ($M_{\odot}$) 
		for the $l=2$ f-mode. The
		colors represent different values of the scalar field
		mass $m_{\phi}$, with the general relativistic limit
		in black. 
	}
	\label{fig:fmode_sly}
\end{figure}

We start our discussion of the results with a detailed
analysis of the properties of the quadrupole f-mode.
The $l=2$ fundamental mode is probably the most interesting
mode as regards to astrophysical scenarios, 
since simulations in General Relativity show, 
that it tends to dominate the ringdown spectrum
after a merger. 

In Figure \ref{fig:fmode_sly} we show the $l=2$ fundamental
mode for several values of the scalar field mass $m_\phi$. 
In the left panel we show the frequency as a function 
of the total mass (in units of solar masses), 
and in the right panel the damping time (in seconds)
as a function of the total mass. 
Different colors symbolize different values of the
parameter $a$, and thus the scalar field mass $m_\phi$,
with purple, cyan, red, orange and green
for $m_{\phi}=1.08$, $0.343$, $0.108$, $0.0343$ 
and $0.0108$ neV, respectively. 
In blue we show the massless case, and 
in black the general relativistic values for
comparison. Note that for $1.08$ neV the values are 
already very close to those of General Relativity, 
while below $m_{\phi}=1$ peV the values do not deviate 
significantly from the blue curve. 
The range of masses considered,
$0.0108 \le m_{\phi} \le 1.08$ neV,
is compatible with current observations and constraints
on the mass of a hypothetical ultra light boson
\citep{Naf:2010zy,Brito:2017zvb}.

Overall we observe that the frequency and the damping time
do not change drastically when going to these STTs, 
finding typical variations within $10\%$ of the 
general relativistic value. This behaviour is reminiscent of
the previous observations for the axial modes
\citep{Blazquez-Salcedo:2018qyy}.
Figure \ref{fig:fmode_sly} (left) shows that a decrease
of the mass of the scalar field leads to a decrease of
the frequency, except for values of the stellar mass 
close to one solar mass, where the frequency rises slightly
from the general relativistic value. Regarding the damping
time, Figure \ref{fig:fmode_sly} (right) shows that the
overall value of $\tau$ decreases the more,
the lighter the scalar field is.


\begin{figure}
	\centering
	\includegraphics[width=.38\textwidth, angle =-90]{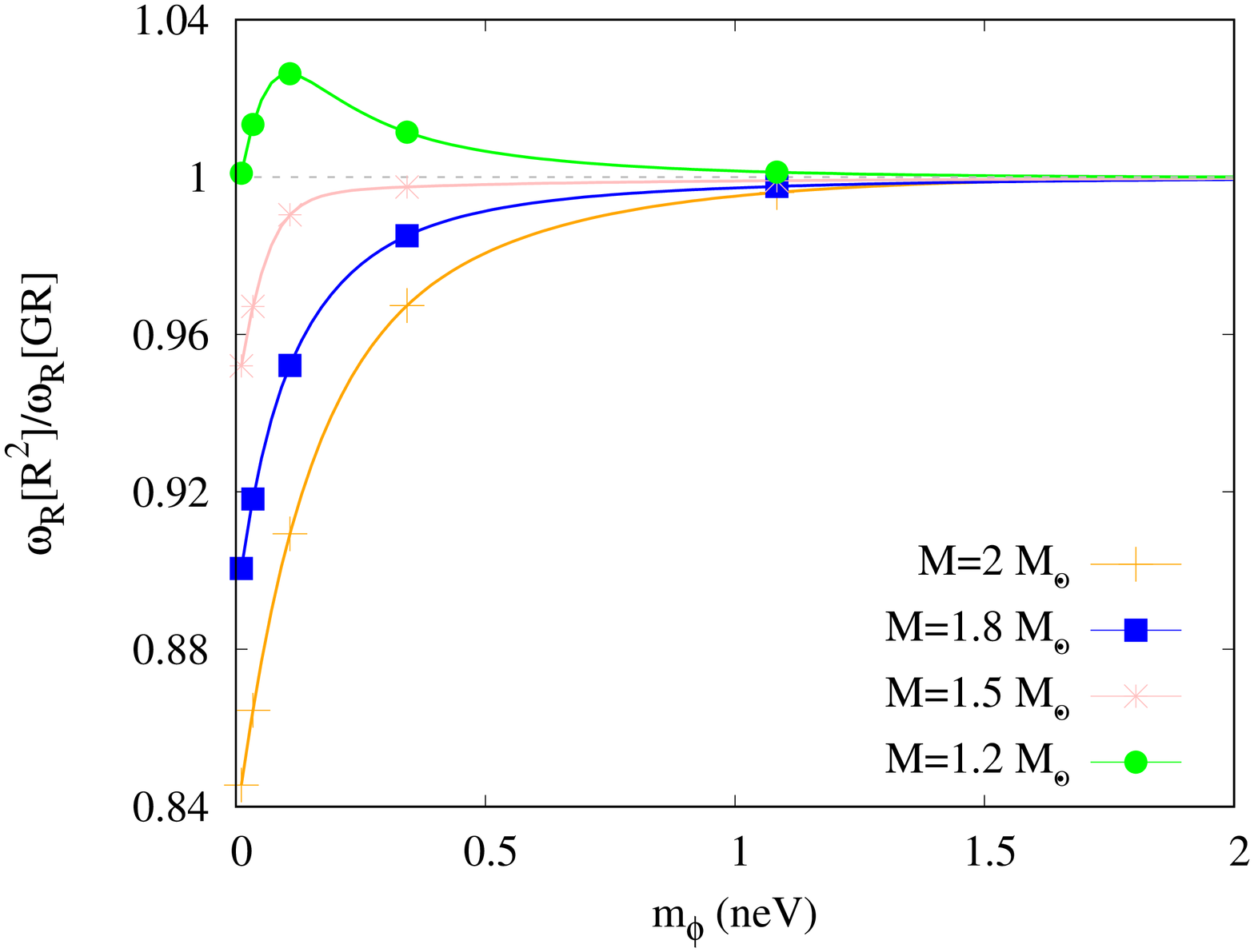}
	\includegraphics[width=.38\textwidth, angle =-90]{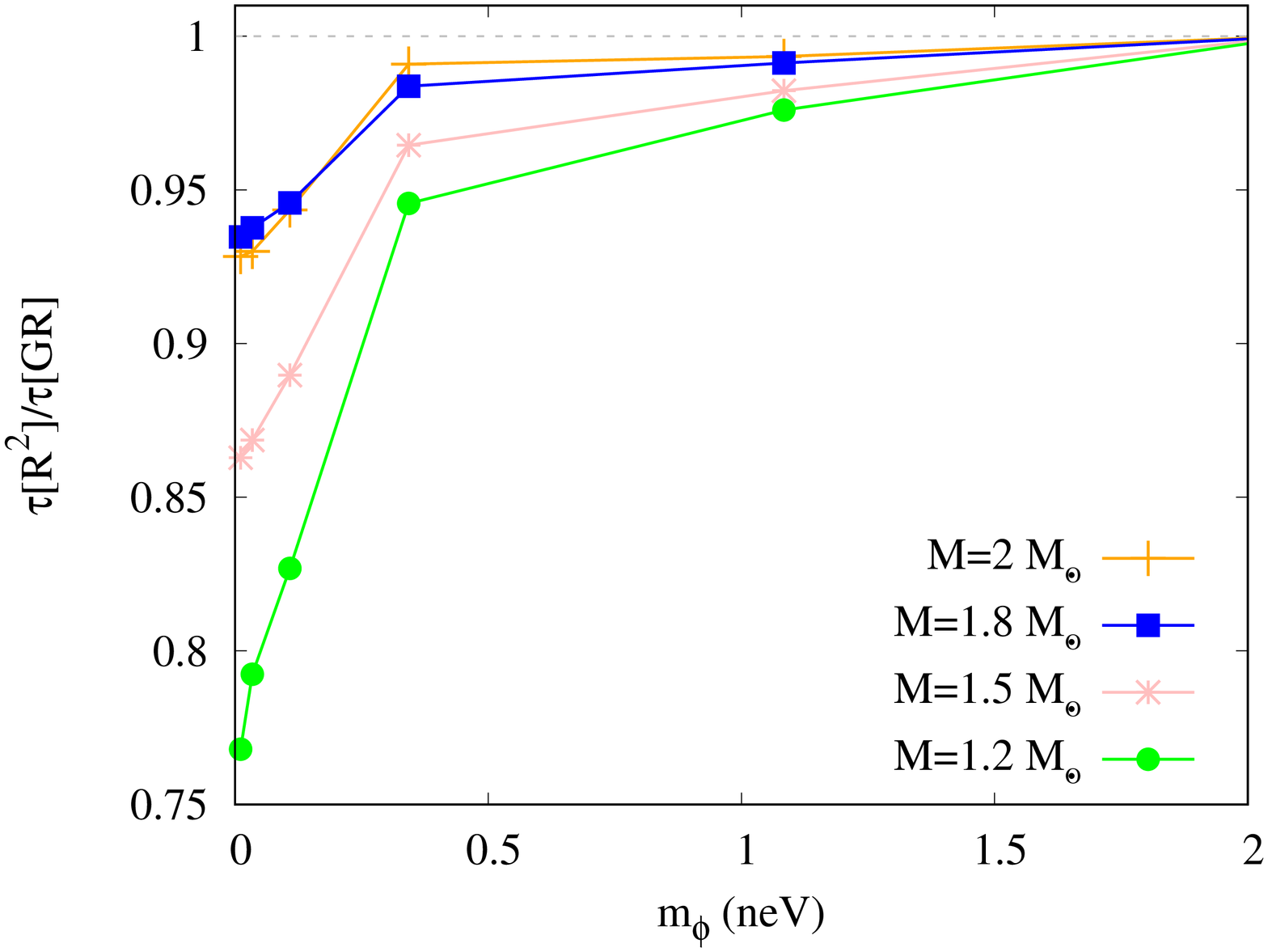}
	\caption{
		Frequency $\omega_R$ in kHz (left) and damping time $\tau$
		in s (right) versus the mass of
		the scalar field $m_{\phi}$ in neV
		for the $l=2$ f-mode. The colors
		represent different values of the total mass $M$ of
		the star. The frequency is normalized in terms of the
		corresponding frequency for a star of given mass in
		General Relativity. This value is obtained
		asymptotically as the mass of the scalar field
		is increased. 
	}
	\label{fig:fmode_fixed_mass}
\end{figure}

To better understand the dependence of the fundamental mode 
on the mass $m_\phi$ of the scalar field, we now fix 
the mass $M$ of the star, while we vary $m_\phi$. 
In Figure \ref{fig:fmode_fixed_mass} we show the $l=2$
fundamental mode as a function of the scalar field mass
$m_\phi$. Each curve corresponds to a family of stars with
fixed value of the mass, with orange, blue, pink and green for
$M=2$, $1.8$, $1.5$, and $1.2 M_{\odot}$, respectively. 
On the left we show the frequency and on the right the
damping time, both normalized to the 
respective value in General Relativity. 
The figure clearly shows that the larger the scalar mass, 
the closer the frequency comes to the general relativistic
value. In fact, the frequency and the damping time deviate
only significantly when the scalar field mass
is below the neV. The shortest damping time occurs for 
massless scalar fields, and the largest deviation occurs 
for not very massive neutron stars with $M=1.2 M_{\odot}$. 
We note that for $M=1.2 M_{\odot}$, the frequency rises 
slightly as the scalar mass is decreased. 
However, in general for sufficiently massive neutron stars, 
the maximum deviation of the frequency appears also for
massless scalar fields, with a reduction of the frequency of
around $10\%$. 

\begin{figure}
	\centering
	\includegraphics[width=.38\textwidth, angle =-90]{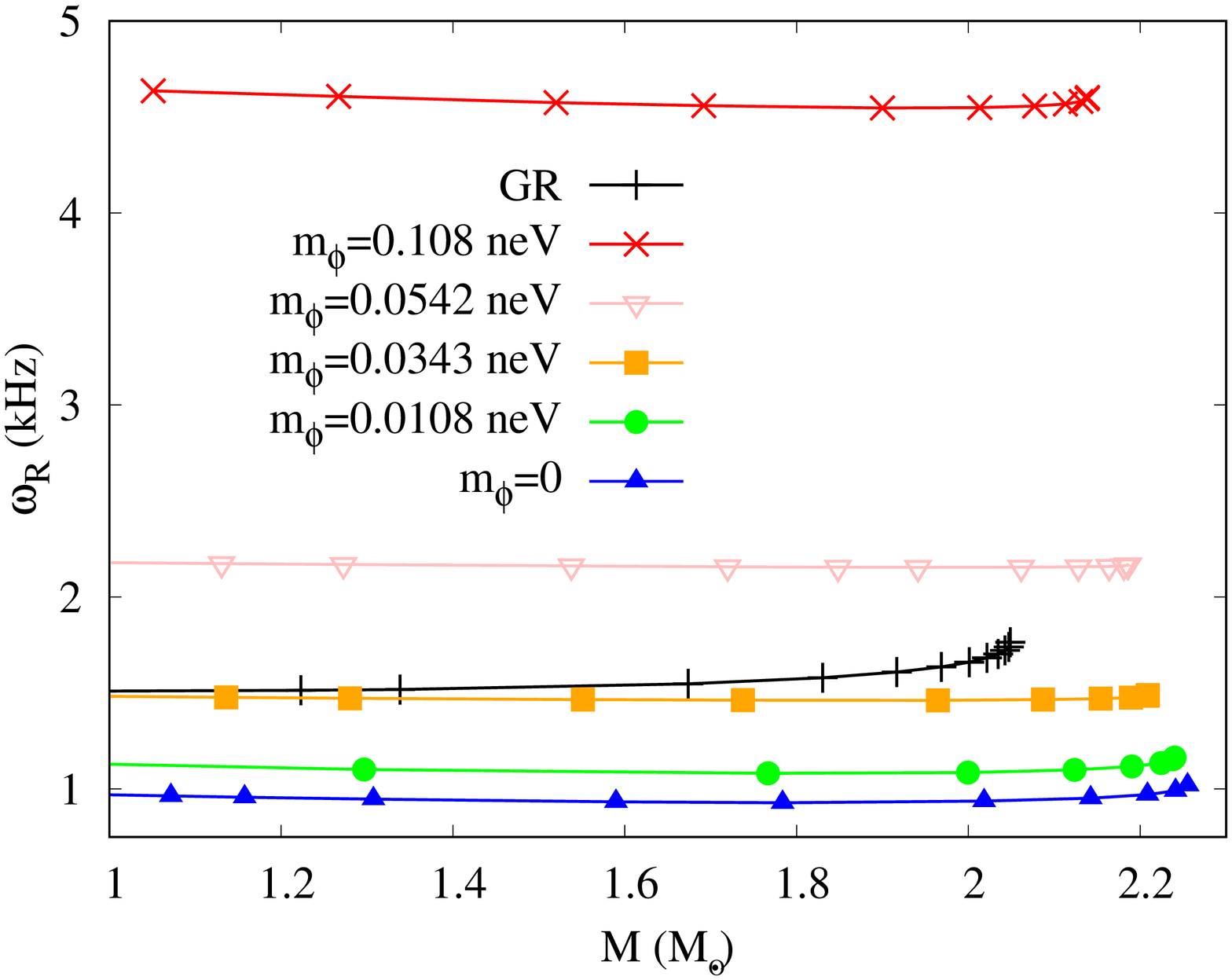}
	\includegraphics[width=.38\textwidth, angle =-90]{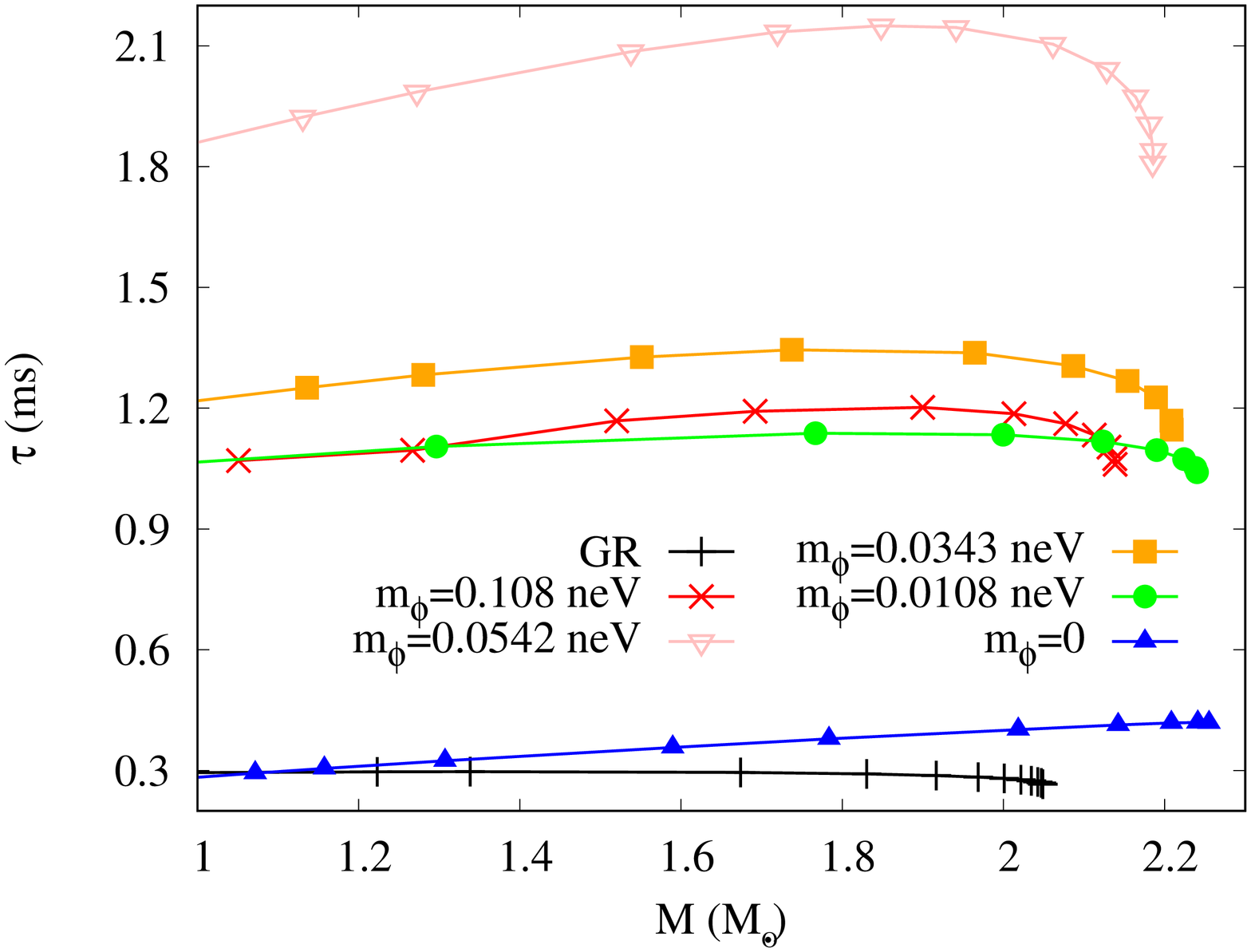}
	\caption{
		Frequency $\omega_R$ in kHz (left) and damping time $\tau$ 
		in ms (right) versus the total
		mass of the neutron star $M$ ($M_{\odot}$) 
		for the $l=2$ $\phi$-mode. The
		colors represent different values of the scalar field
		mass $m_{\phi}$, with the general relativistic limit
		in black.
	}
	\label{fig:s2mode_fixed_mass}
\end{figure}

The presence of the scalar field allows for
an additional type of $l=2$ mode. 
In General Relativity this mode would
correspond to the mode of an independent minimally coupled
scalar field in the background of the neutron star,
and therefore not be of much interest.
In STTs, however, the equations for the scalar field 
perturbation and the matter and metric perturbations
are coupled for polar modes. Therefore this new type
of mode is necessarily present, and is dubbed
$l=2$ $\phi$-mode, distinguishing it from the previously
discussed $l=2$ f-mode.

In Figure \ref{fig:s2mode_fixed_mass} we show the $l=2$
$\phi$-mode as a function of the 
mass $M$ of the neutron star for several values of the
scalar field mass $m_\phi$ as well as for General Relativity.
Interestingly, there is very little dependence
of the frequency and the damping time on
the neutron star mass.
Moreover, for most of the relevant mass range
of the scalar field in these STTs the frequencies
of these $\phi$-modes are lower than the frequencies
of the corresponding f-modes. In contrast, the
damping times of the $\phi$-modes (of the order ms) are much shorter
than the damping times of the corresponding f-modes.
In a ringdown spectrum the $\phi$-modes will therefore 
fast decay, leaving the f-modes to dominate the spectrum.

Let us also mention here that, although we have not made a systematic study of the $l=2$ excited modes, they also exist in the STTs. Our numerical results indicate that the overtones for both the f-mode and the $\phi$-mode always possess larger frequencies and shorter damping times than the corresponding ground states.

\vspace{0.5cm}

\subsection{Radial modes}

\begin{figure}[t!]
	\centering
	\includegraphics[width=.38\textwidth, angle =-90]{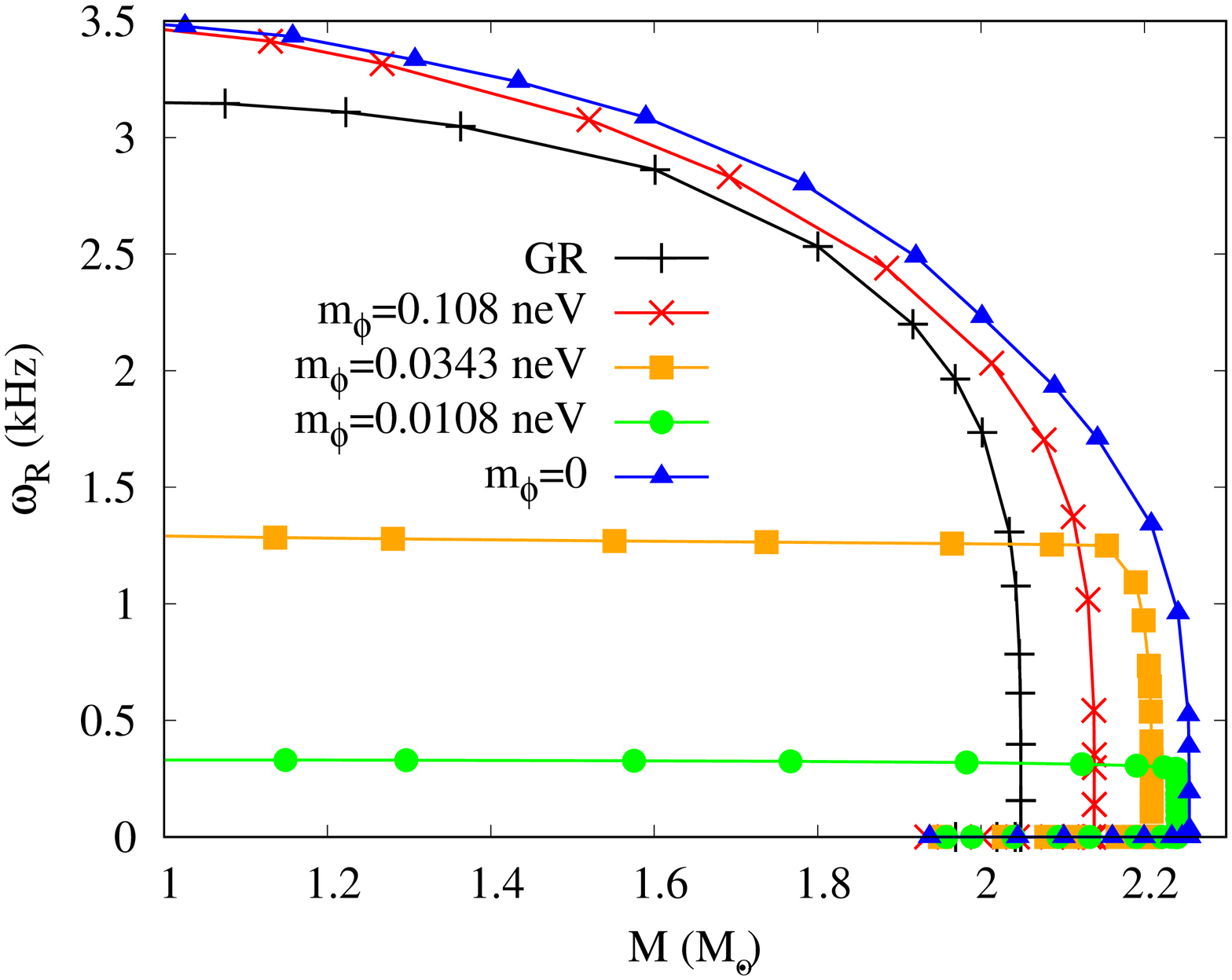}
	\includegraphics[width=.38\textwidth, angle =-90]{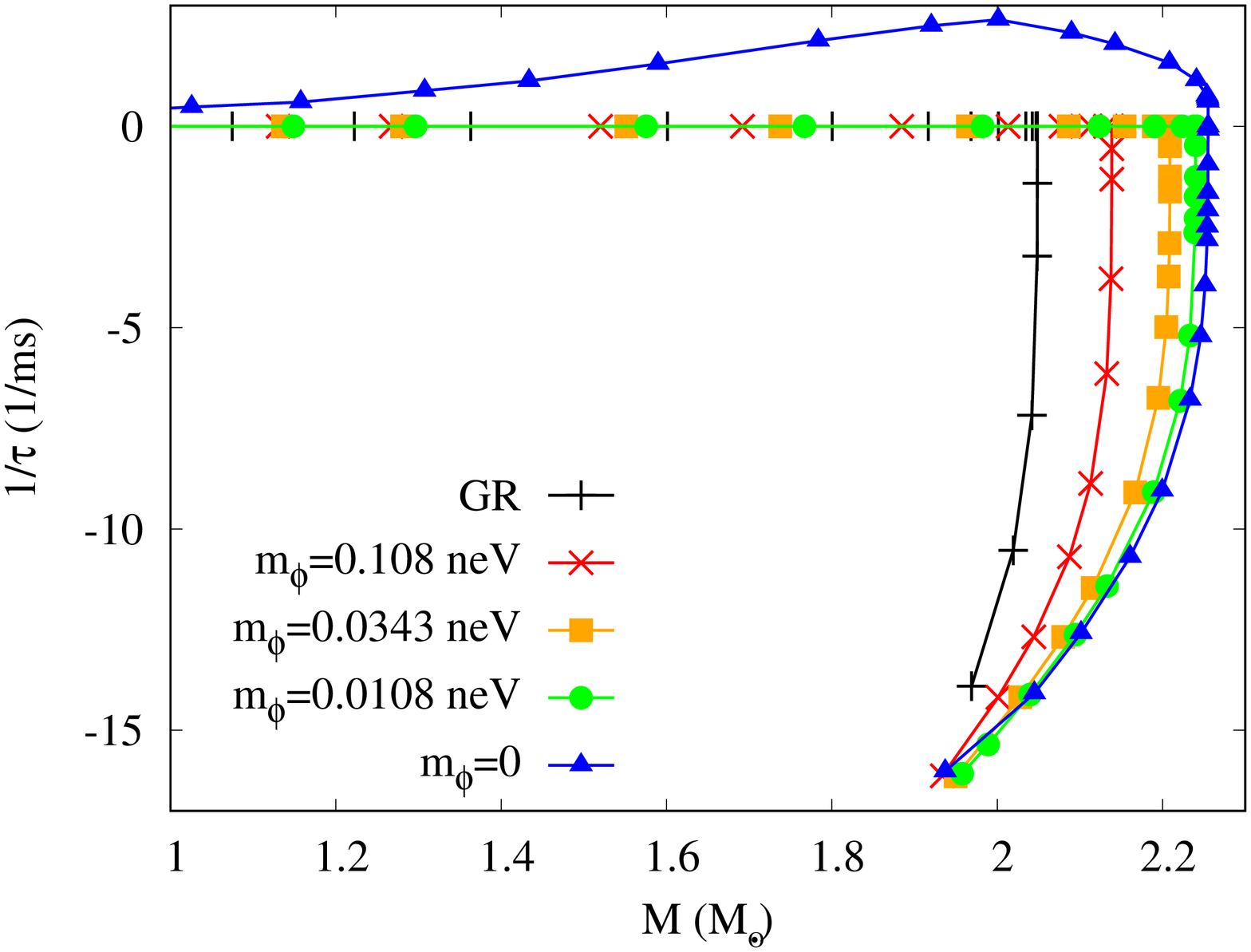}
	\caption{
		Frequency $\omega_R$ in kHz (left) and inverse 
		instability timescale $\omega_I=1/\tau$
		in 1/ms (right) versus the total mass of
		the neutron star $M$ ($M_{\odot}$) 
		for the $l=0$ F-mode.
		 The colors represent different values of the scalar
		 field mass $m_\phi$, with the general relativistic limit in black.
	}
	\label{fig:radial_F_modes_sly}
\end{figure}

Gravitational monopole radiation does not exist in GR. 
Hence, the ringdown phase in astrophysically realistic 
scenarios is expected to be dominated by the $l=2$ modes. 
In neutron stars the dominant mode would be the f-mode,
discussed above.
However, the additional scalar degree of freedom present
in STTs implies that,
in astrophysical scenarios, apart from the $l=2$ f-mode,
also the radial $l=0$ modes could play an important role,
when the neutron stars carry scalar hair.

The analysis of the $l=0$ modes in STTs reveals indeed 
a rich spectrum, consisting of two families of modes, 
named again according to their general relativistic limits.
First, there are the modes that reduce to the oscillations 
of the nuclear matter in the general relativistic limit.
These are the fundamental pressure-led mode, the F-mode,
and its excitations, the H$_1$-mode, the H$_2$-mode, etc. 
Second, there are the modes that reduce to oscillations 
of an independent minimally coupled
scalar field in the general relativistic limit.
These are the $l=0$ $\phi$-modes.

As discussed above, in General Relativity the equations 
for these two types of modes are completely decoupled.
The physically interesting modes in General Relativity
are the pressure-led modes. They represent \textit{normal modes},
since $l=0$ modes are confined to the interior of the stars here.
The fundamental pressure-led mode, the F-mode,
is of particular relevance, since it shows,
that neutron stars become radially unstable 
beyond their maximum mass.

For the F mode the eigenvalue $\omega$ 
is a positive real number,
as long as the mass $M$ of the neutron star
increases with increasing central pressure,
and the star is radially stable.
Then $\omega$ becomes zero as the maximal neutron star mass 
is reached. Thus here a zero mode is encountered 
where the star is only marginally stable.
Beyond the maximum mass of the star, $\omega$ becomes 
purely imaginary with $\omega_I$ negative,
i.e., the star becomes radially unstable.

The $\phi$-modes, on the other hand, could in principle
propagate outside the star, in case some external scalar
field were minimally coupled to General Relativity,
and a quasinormal mode analysis does yield a spectrum
of such damped radial modes.
However, the motivation for the presence of such a field
in the environment of a neutron star would be typically
lacking.

In STTs the picture changes almost completely. 
Only the instability of the stars beyond the maximum mass
is still revealed by a purely imaginary eigenvalue with
negative $\omega_I$.
Most importantly, however, scalar radiation is a natural effect 
in STTs, since the
scalar field is a relevant gravitational degree of freedom,
that is intimately coupled with the tensor degrees of freedom.
Consequently, the stable normal modes from General Relativity
now turn into propagating and thus quasinormal modes 
(see e.g., the toy model studied in \citep{Kokkotas:1986gd}): 
the presence of the gravitational scalar field now
allows for these $l=0$ modes to propagate outside the star. 

We exhibit in Fig.~\ref{fig:radial_F_modes_sly}
the frequency $\omega_R$ in kHz (left) and the 
imaginary part $\omega_I$ in 1/ms (right) for the F-mode
versus the total mass $M$ in solar masses $M_{\odot}$
for the neutron stars. As before, the colors represent 
different values of the scalar field mass $m_\phi$, 
and the general relativistic limit is shown in black.
On the unstable neutron star branches beyond the 
maximum mass $\omega_I=1/\tau$ represents the
inverse instability timescale.

On the stable neutron star branches, in contrast,
the very small positive values of $\omega_I$
indicate the inverse damping time. 
While numerical inaccuracy does not allow us to
precisely extract these values, 
the calculations indicate that the damping time
is on the order of $\sim 10^5$ years or larger.
This means that these F-modes are ultra long lived
\citep{Blazquez-Salcedo:2020ibb}.
In fact, these modes exist up to $m_\phi \to \infty$,
where they turn into the normal modes of the respective
general relativistic stars.
Interestingly, long lived scalar radiation 
was also detected in core collapse processes 
in massive STTs \citep{Sperhake:2017itk,Rosca-Mead:2020ehn,Rosca-Mead:2020cyo,Rosca-Mead:2020bzt}.

\begin{figure}
	\centering
	\includegraphics[width=.38\textwidth, angle =-90]{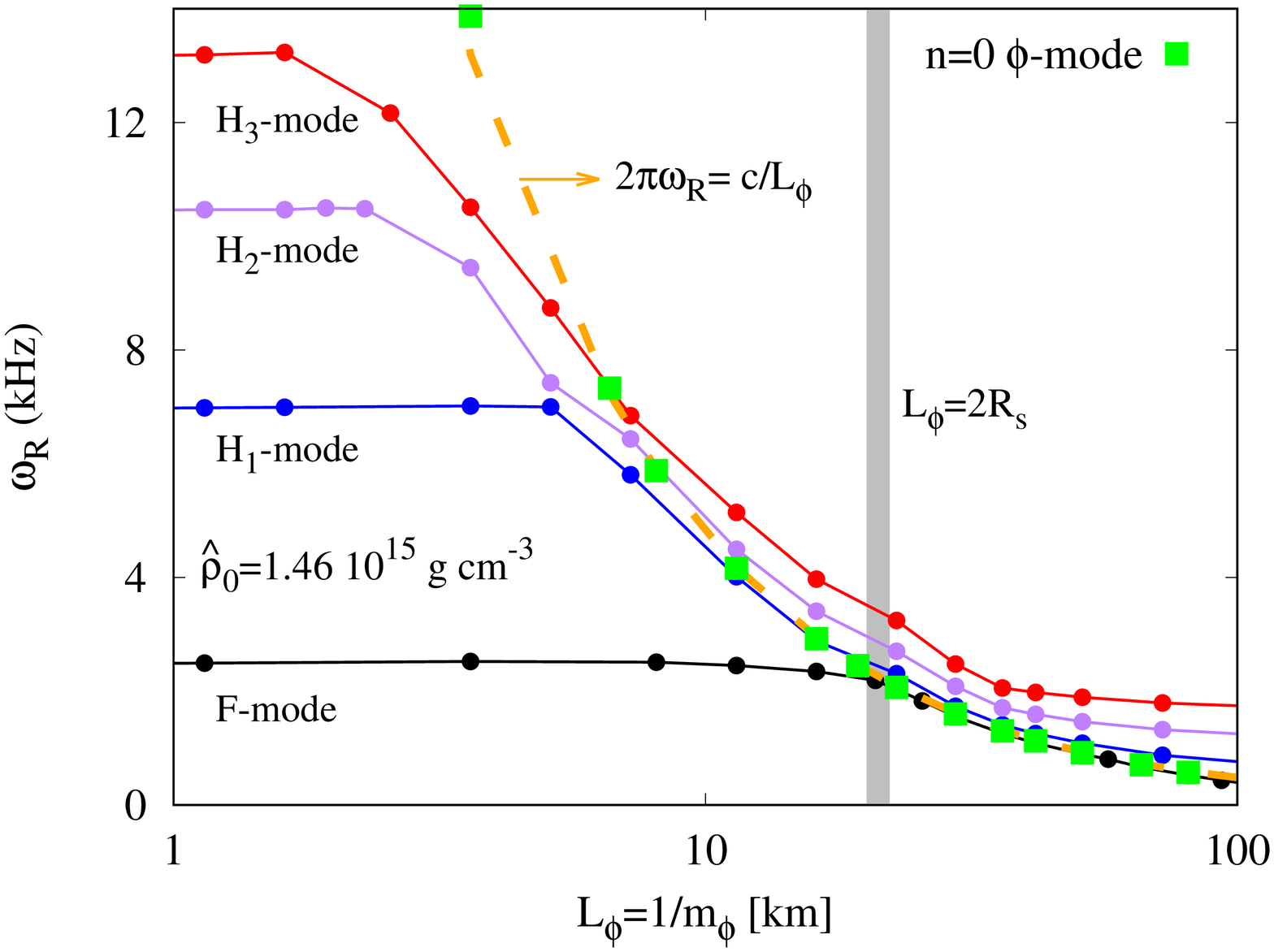}
	\includegraphics[width=.38\textwidth, angle =-90]{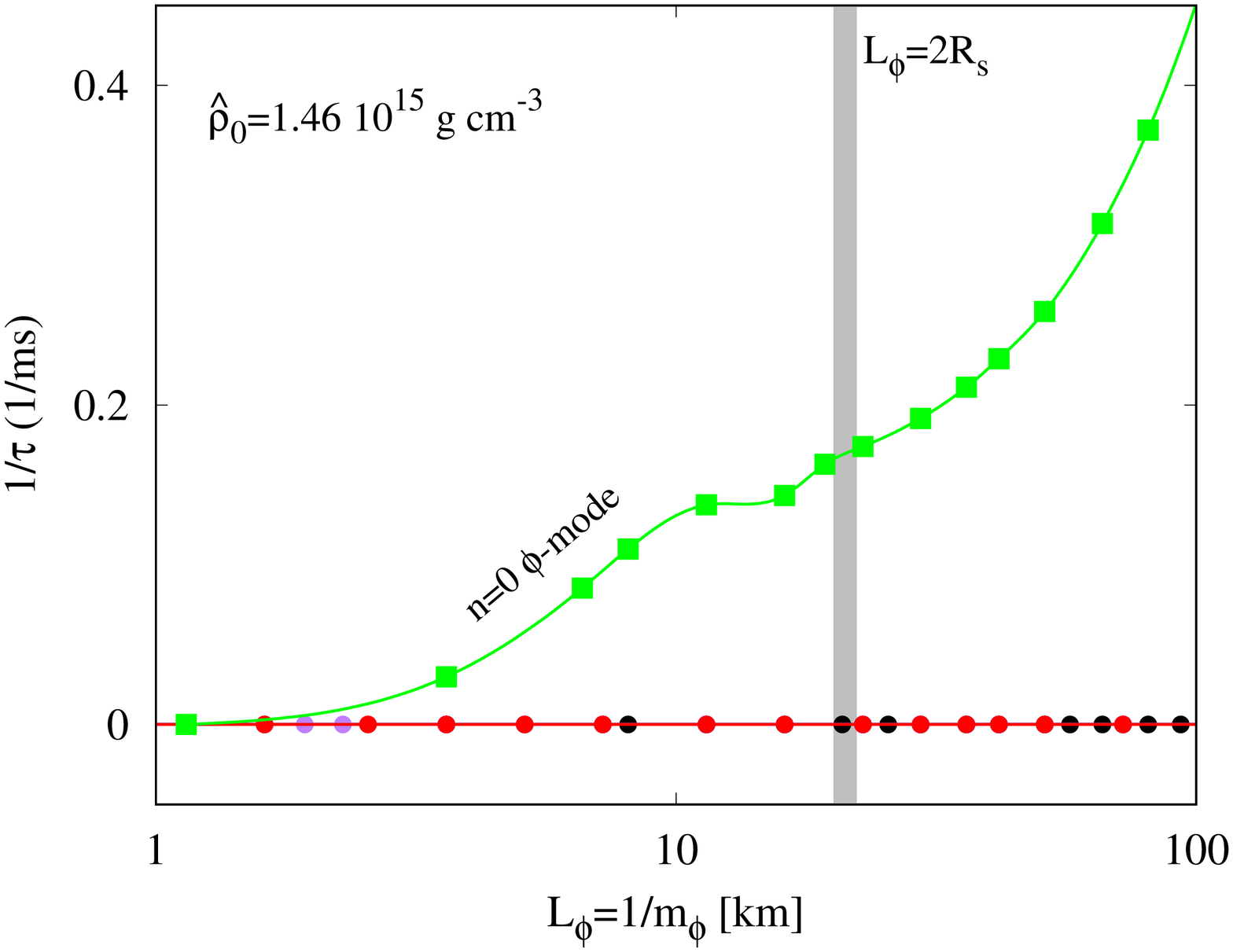}
	\caption{
	Frequency $\omega_R$ in kHz (left) and inverse 
		damping time $\omega_I=1/\tau$
		in 1/ms (right) versus the Compton wavelength
		$L_\phi=1/m_\phi$ in km
		for the $l=0$ F-, H$_1$-, H$_2$- and H$_3$-modes
		at fixed central density $\hat{{\rho}}_0$. 
		For comparison, the inverse of the Compton
		wavelength $L_\phi$ is shown in orange (left) and
		the size of the neutron star $L_\phi=2R_s$ in grey.
	}
	\label{fig:excited_H_mode_sly}
\end{figure}

In Figure \ref{fig:excited_H_mode_sly} we show 
the frequency $\omega_R$ in kHz (left) and 
the inverse damping time $\omega_I=1/\tau$ in 1/ms (right) 
versus the Compton wavelength $L_\phi=1/m_\phi$ in km
for the F-mode and the excited H-modes, H$_1$ -- H$_3$,
for a fixed value of the central density
on the stable neutron star branch.

We note, that also for the excited modes
the spectrum is qualitatively similar to the GR case.
But again the fundamental difference is that the normal modes 
of General Relativity turn into quasinormal modes
in the STTs, that are allowed to propagate outside the star, 
since the pressure-led modes induce also oscillations 
in the scalar field, which itself is coupled to the metric 
functions in the field equations.
Moreover, the excited H-modes are ultra long lived, as well.
As an aside we note, that 
in STTs with spontaneously scalarized neutron stars, analogous
ultra long lived quasinormal modes have not been observed
\citep{Mendes:2018qwo}.

For comparison Figure \ref{fig:excited_H_mode_sly} (left)
also shows the frequency which simply corresponds to
the inverse of the Compton wavelength of the scalar field,
\begin{equation}
    2\pi \omega_R = c/L_{\phi} ,
\end{equation}
and the size of the neutron star
\begin{equation}
  L_{\phi} = 2 R_s , 
\end{equation}
where $R_s$ represents the radius of the family of stars.
The figure shows, that for small values of the
Compton wavelength $L_\phi$ the frequencies 
$\omega_R$ of the F-mode and the excited H-modes
assume almost constant values, that increase with
increasing excitation.

The (almost) $m_\phi$-independence ends for the fundamental F-mode
when $L_\phi$ reaches the size of the star, and,
analogously, for the excited H-modes, when appropriate
fractions of the size of the star are reached.
For the F-mode this happens when the Compton wavelength
reaches $L_\phi=2R_s$ with $m_{\phi}=0.052$ neV,
and for the first three H-modes for
$m_{\phi}=0.072, 0.16, 0.19$ neV, respectively.
For larger values of the Compton wavelength $L_\phi$
the frequencies decrease according to
$ \omega_R = 1/ 2 \pi L_\phi$,
thus the frequencies simply follow the scale 
given by the mass of the scalar field, as
depicted by the orange curve in the left figure.

Figure \ref{fig:excited_H_mode_sly} (right) shows that
the inverse decay times $1/\tau$ of the excited modes 
are indeed also very small. When estimating numerically
the damping distances for these modes one finds that
they should be equal or larger than $\sim 10^5$ ly.
This corresponds to the size of a large
galaxy like the Milky Way, and possibly even larger.

\begin{figure}
	\centering
	\includegraphics[width=.38\textwidth, angle =-90]{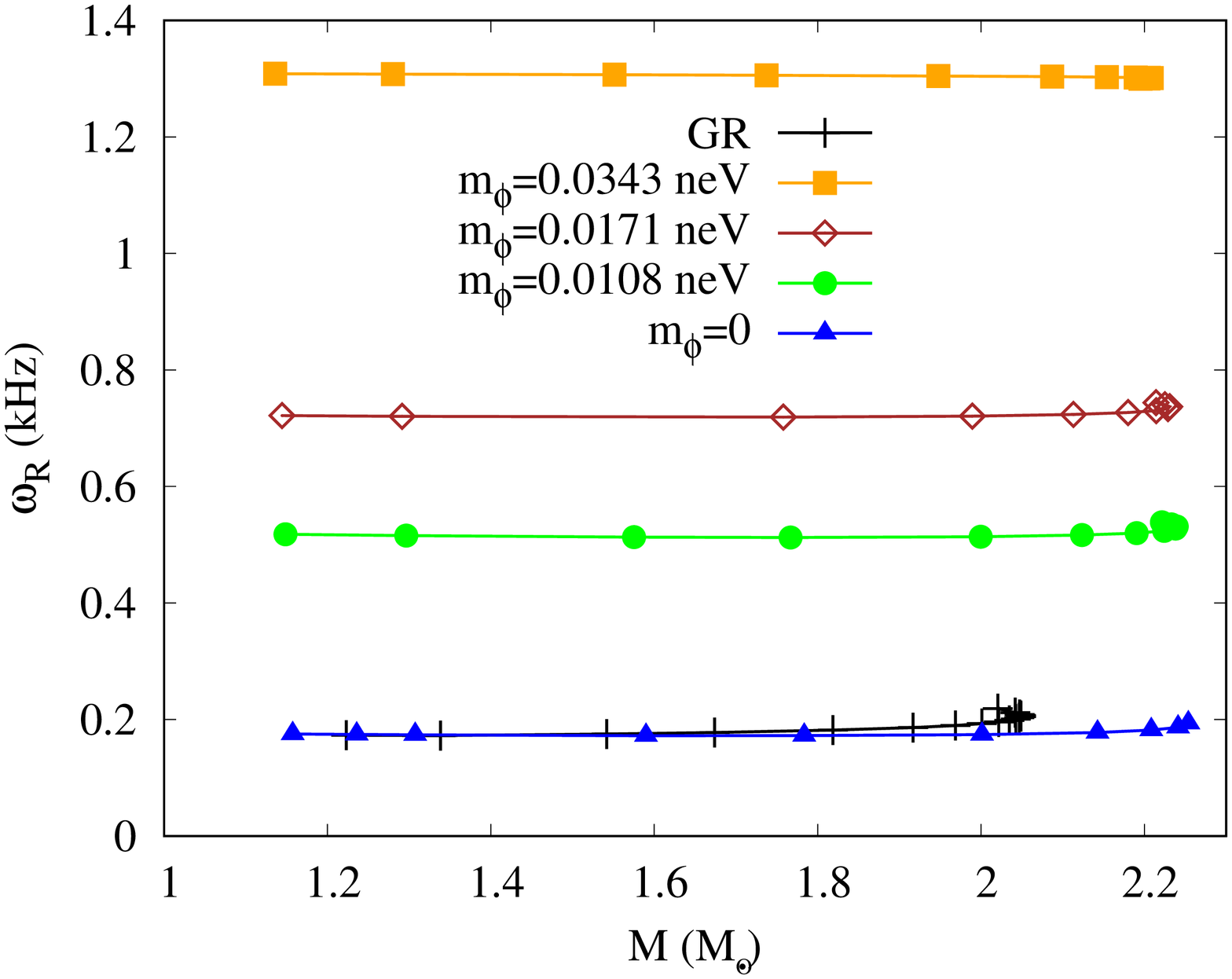}
	\includegraphics[width=.38\textwidth, angle =-90]{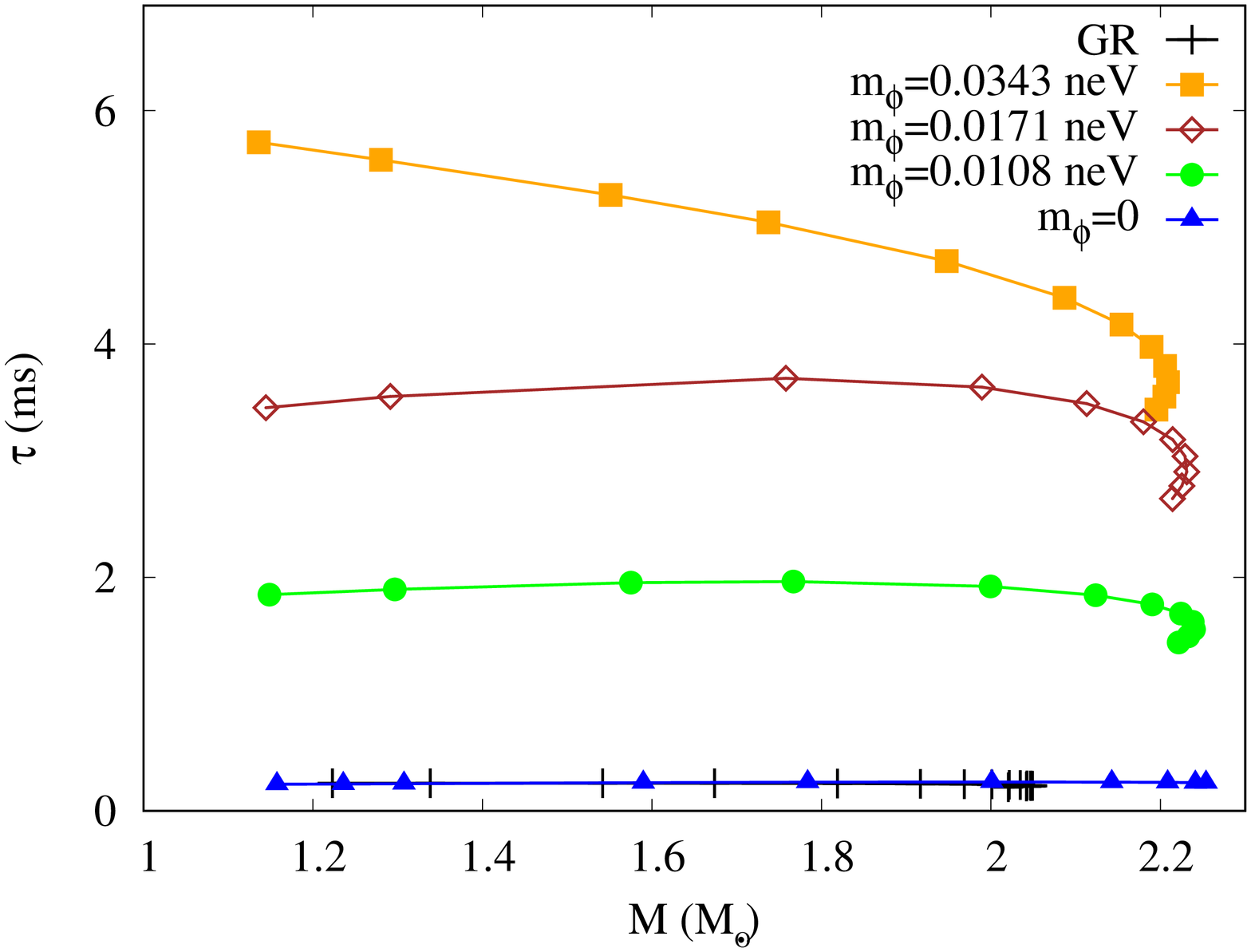}
	\caption{
		Frequency $\omega_R$ in kHz (left) and damping time $\tau$
		in ms (right) versus the total
		mass of the neutron star $M$ ($M_{\odot}$) 
		for the $l=0$ $\phi$-mode. The
		colors represent different values of the scalar field
		mass $m_{\phi}$, with the general relativistic limit
		in black. 
	}
	\label{fig:s0mode_fixed_mass}
\end{figure}

As discussed above, in addition to the pressure-led $l=0$ modes
there are also the scalar-led $l=0$ modes, the $\phi$-modes.
Figure \ref{fig:excited_H_mode_sly} also shows the fundamental
$l=0$ $\phi$-mode. Its frequency $\omega_R$ follows always the 
scalar mass $m_\phi$, as seen by the overlap of the green dots
($\phi$-mode) and the orange curve (left). 
However, its imaginary part $\omega_I=1/\tau$ increases rapidly
with increasing Compton wavelength $L_\phi$, showing 
that these modes possess much shorter lifetimes.

In Figure \ref{fig:s0mode_fixed_mass} we exhibit the
frequency $\omega_R$ (kHz) (left) and the
damping time $\tau$ (ms) (right)
versus the total mass of the neutron star $M$ ($M_{\odot}$)
for the $l=0$ $\phi$-mode. %
As already seen for the $l=2$ $\phi$-mode,
the frequency $\omega_R$ changes only very little with the
neutron star mass.
The damping time shows so little dependence only
in the limiting case of General Relativity, and
for the massless scalar field.

As a comment, let us note here that scalar-led $l=0$ modes appear also in simulations of oscillating and collapsing neutron stars in chamaleon theories, in addition to the fluid modes \citep{Dima:2021pwx}.

\vspace{0.5cm}

\subsection{Comparison of potentials}

\begin{figure}
	\centering
	\includegraphics[width=.38\textwidth, angle =-90]{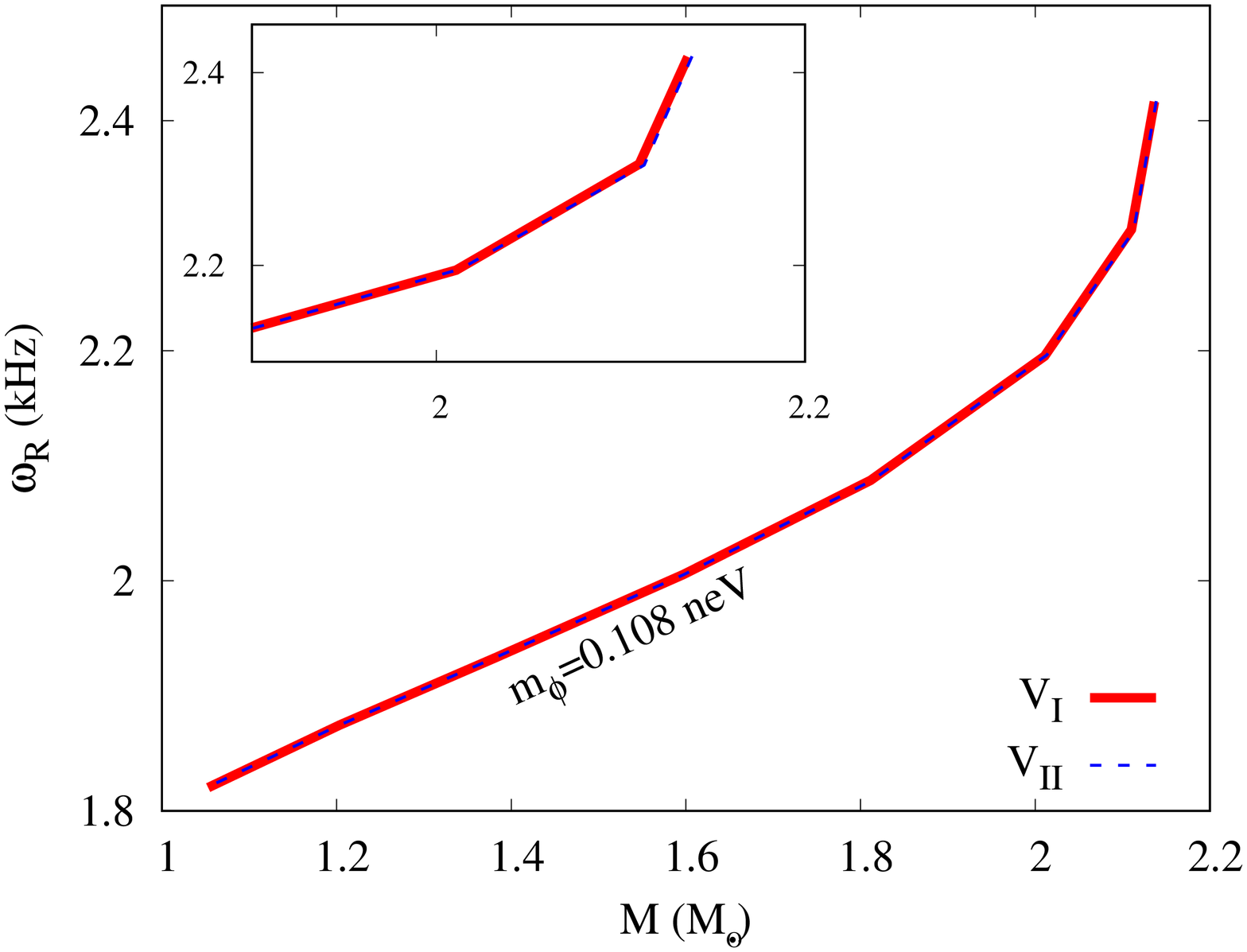}
	\includegraphics[width=.38\textwidth, angle =-90]{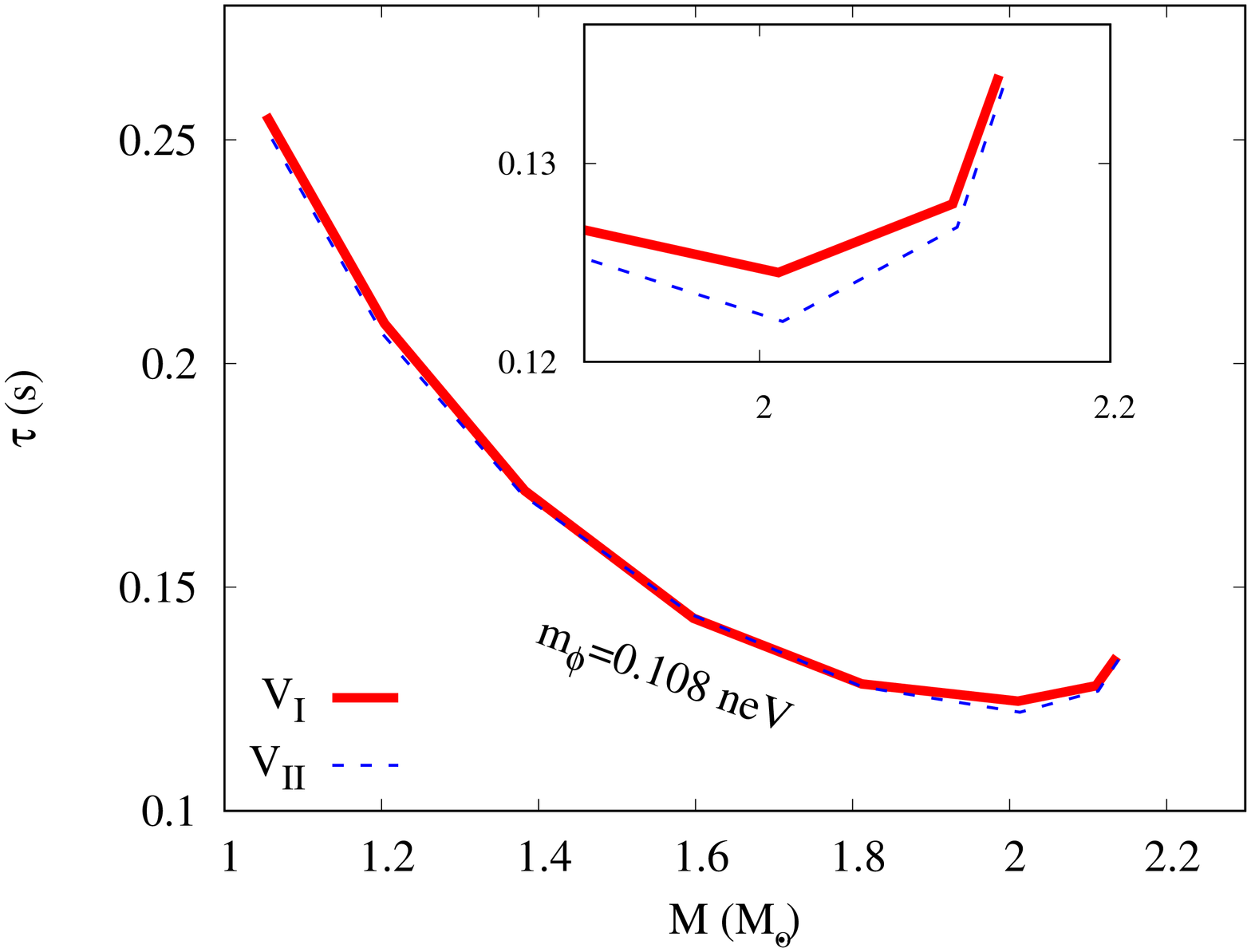}
	\caption{
		Frequency $\omega_R$ in kHz (left) and damping time $\tau$
		in s (right) versus the total
		mass of the neutron star $M$ ($M_{\odot}$) 
		for the $l=2$ f-mode. The colors represent the two
		potentials $V_I$ and $V_{II}$.
	}
	\label{fig_comparison_V1}
\end{figure}

\begin{figure}
	\centering
	\includegraphics[width=.38\textwidth, angle =-90]{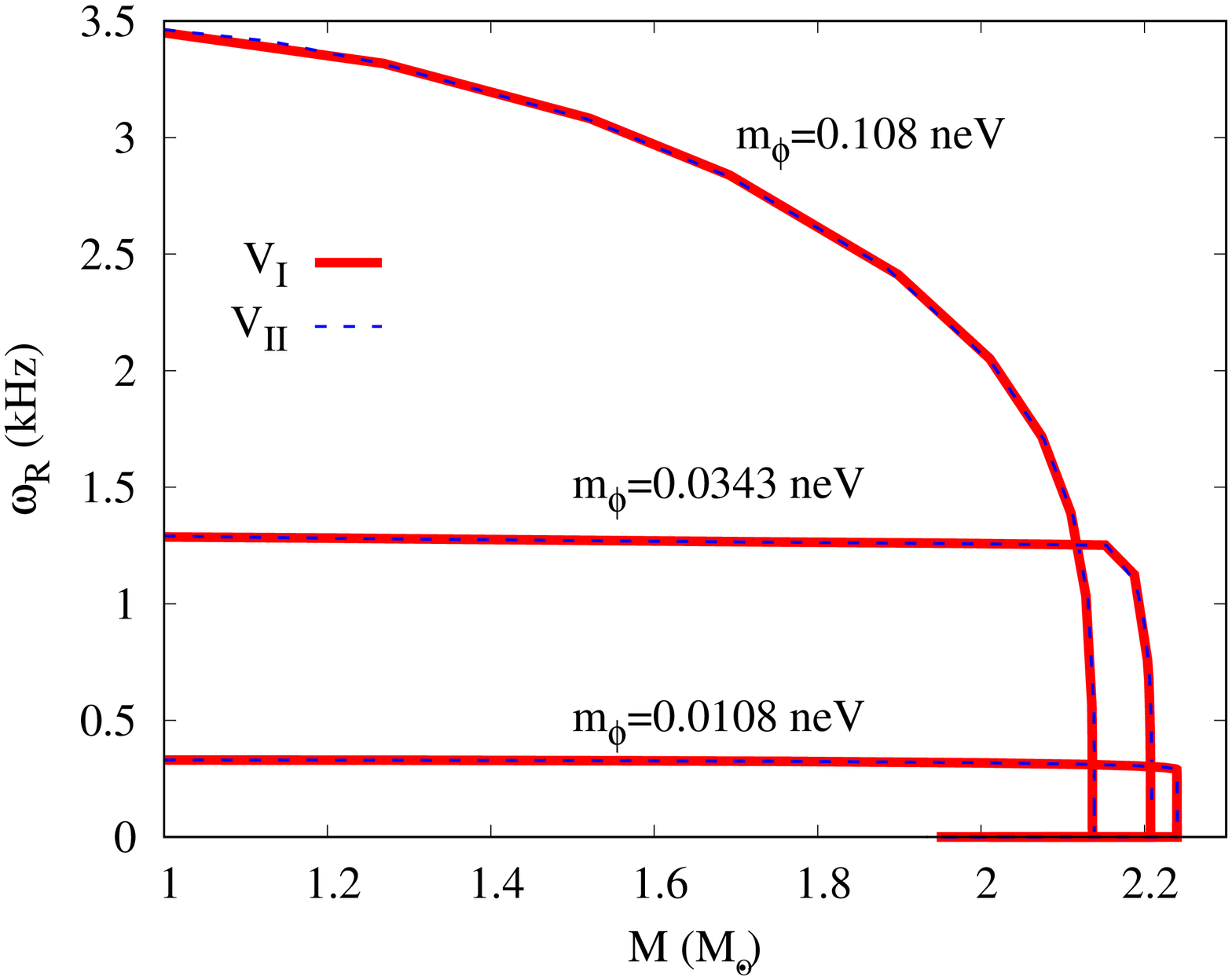}
	\includegraphics[width=.38\textwidth, angle =-90]{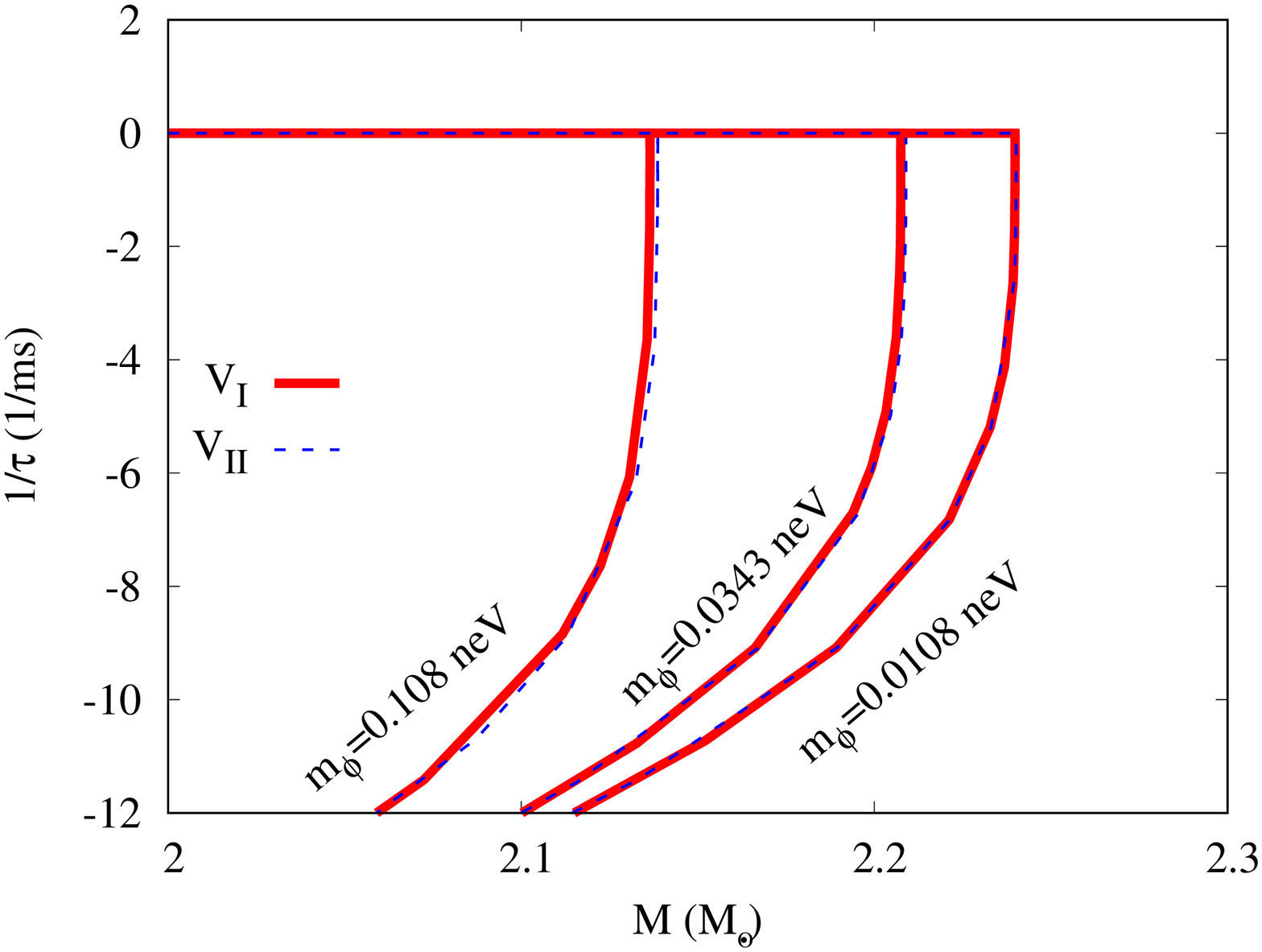}
	\caption{
	Frequency $\omega_R$ in kHz (left) and inverse 
		instability timescale $\omega_I=1/\tau$
		in 1/ms (right) versus the total
		mass of the neutron star $M$ ($M_{\odot}$) 
		for the $l=0$ F-mode. The colors represent the two
		potentials $V_I$ and $V_{II}$.
	}
	\label{fig_comparison_V2}
\end{figure}

\begin{figure}
	\centering
	\includegraphics[width=.38\textwidth, angle =-90]{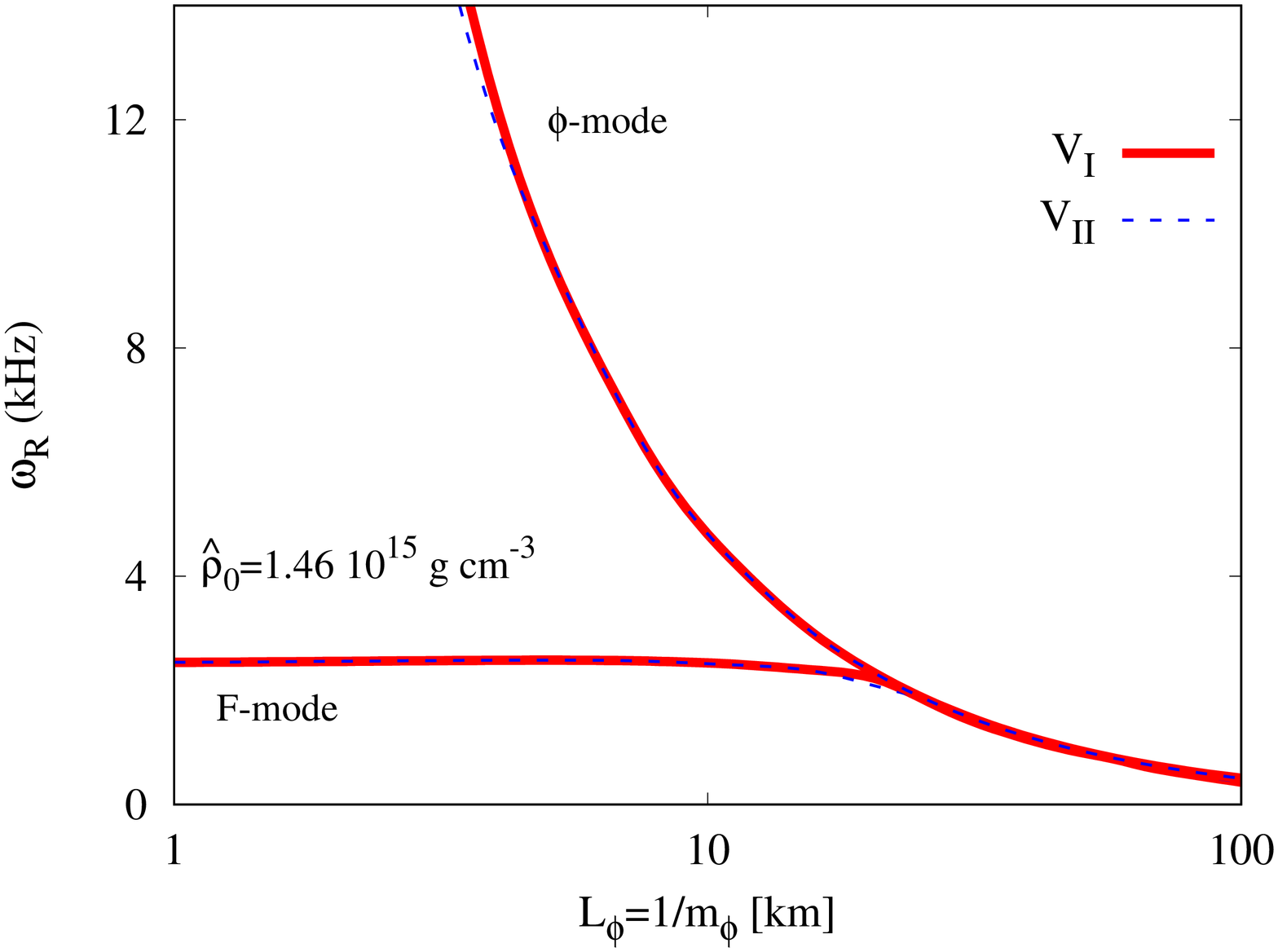}
	\includegraphics[width=.38\textwidth, angle =-90]{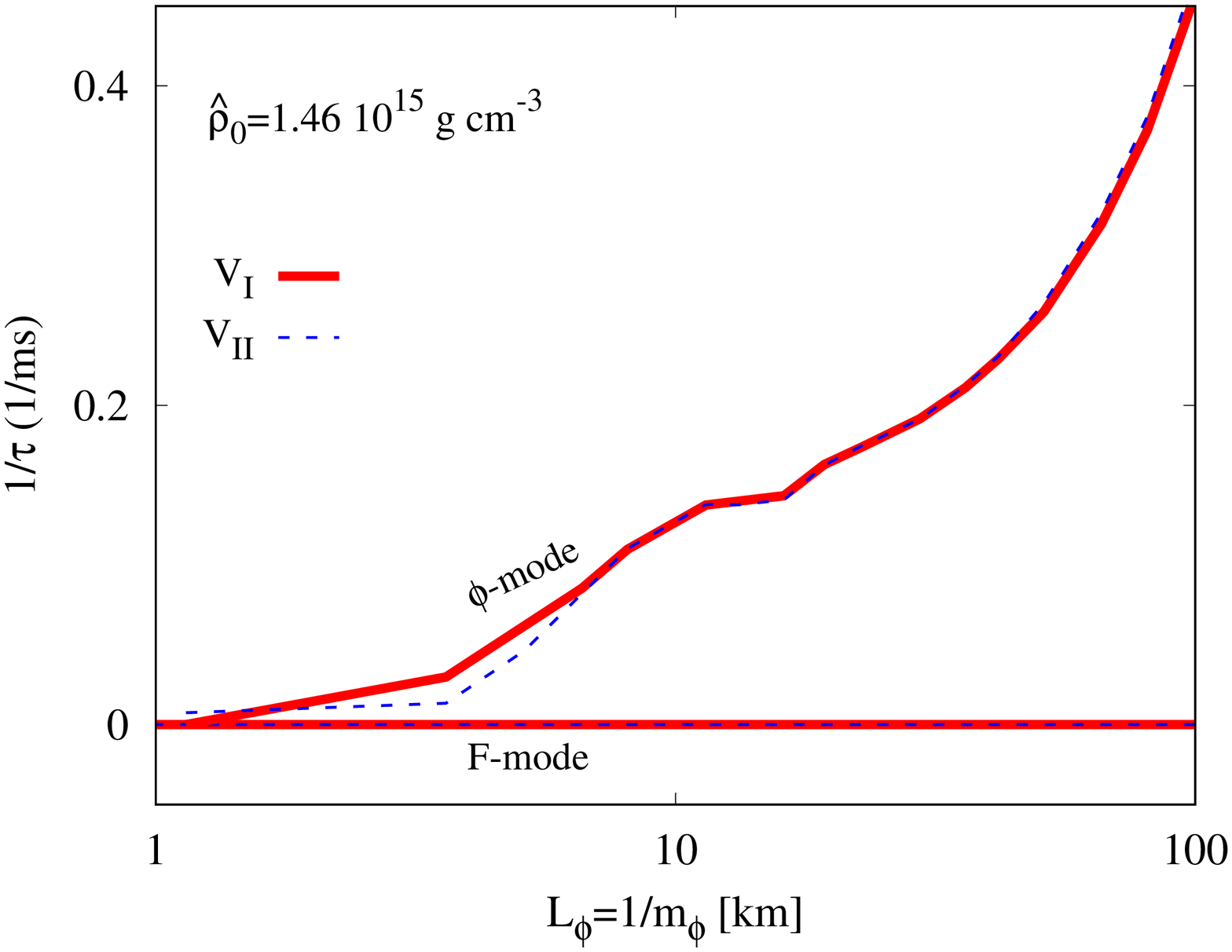}
	\caption{
	Frequency $\omega_R$ in kHz (left) and 
	and inverse damping time $\omega_I=1/\tau$
		in 1/ms (right) versus the Compton wavelength
		$L_\phi=1/m_\phi$ in km
		for the $l=0$ F-mode and $l=0$ $\phi$-mode 
		at fixed central density $\hat \rho_0$. 
		The colors represent the two
		potentials $V_I$ and $V_{II}$.
	}
	\label{fig_comparison_V3}
\end{figure}

We finally briefly address the effect of
the two different potentials,
$V_I$ and $V_{II}$ (see Eqs.~(\ref{pots})),
recalling that $V_I$ represents a STT with
a mass term only, whereas $V_{II}$ 
is the potential derived from $R^2$ gravity.
In Figure \ref{fig_comparison_V1}
we exhibit a comparison of the $l=2$ f-mode
for both potentials and scalar mass
$m_\phi=0.108$ neV. The insets highlight 
the differences.

In Figures \ref{fig_comparison_V2} and \ref{fig_comparison_V3}
we exhibit a comparison of the radial modes.
Figure \ref{fig_comparison_V2} shows the $l=0$ F-mode
versus the neutron star mass for several values of
the scalar field mass.
Figure \ref{fig_comparison_V3} shows the $l=0$ F-mode
and the $l=0$ $\phi$-mode versus the Compton wavelength
of the scalar field for fixed central density.

From the figures we conclude, that
for the range of masses $m_\phi$ of the scalar field
that are of interest \citep{Naf:2010zy,Brito:2017zvb}, 
i.e., $10^{-9} \gtrsim m_\phi \gtrsim 10^{-13}$ eV,
there is very little deviation between the 
quasinormal modes in these two theories.
Deviations are most noticeable for the most compact configurations, close to the maximum mass, and in particular in the damping times.

\section{Conclusions} \label{sec_conclusions}

In this paper we have studied polar quasinormal 
modes of neutron stars in two STTs, one possessing a simple scalar
field mass term and one corresponding to $R^2$ gravity. 
We have presented the set of equations and the boundary conditions 
necessary to obtain these modes. 
The scalar field mass $m_\phi$ leads to a dispersion relation 
between the frequency of the spacetime oscillations and the scalar field oscillations as infinity is asymptotically approached.

We have analyzed the quasinormal modes for the SLy 
equation of state, a realistic equation of state
that yields for static neutron stars in General Relativity 
a maximum mass slightly above two solar masses.
In the STTs studied, the value of the neutron star maximum mass 
increases with decreasing scalar field mass,
reaching about 2.25 solar masses. For the scalar field
we have covered, in particular,
the physically interesting mass range 
\citep{Naf:2010zy,Brito:2017zvb}.

We have investigated the $l=2$ modes, starting with
the fundamental f-mode, and studied the effect of the mass of
the gravitational scalar field of the STTs. Moreover, we have
studied the $l=2$ $\phi$-modes, which represent a set of
additional quadrupole radiation modes, present only in STTs.
While they depend strongly on the scalar field mass,
they depend only weakly on the neutron star mass.

In General Relativity, the quadrupole 
quasinormal modes represent the lowest multipole modes, 
that can propagate gravitational radiation.
In STTs there is, of course, additional gravitational radiation, 
corresponding to $l < 2$ modes.
Here we have analyzed the $l=0$ quasinormal modes 
present in these STTs. In fact, the radial perturbations 
lead to a rich spectrum of modes.

We have shown that the pressure normal modes of General Relativity,
the fundamental F-mode and its excitations,
turn into propagating quasinormal modes in these STTs.
For the observationally relevant range of the scalar mass
these modes are ultra long lived. Thus the 
damping distance of the corresponding
gravitational radiation corresponds to large distances
of $10^5$ or more light years. %

When considering the dependence of these F-modes 
and the excited H-modes on the Compton wavelength
of the scalar field, one finds almost constant
values for their frequencies as long as the Compton
wavelength is smaller than the size of the star
(F-mode) or appropriate fractions of it (H-modes).
Beyond these points, the frequencies decay with the
inverse of the Compton wavelength.

Besides these pressure-led modes, there are also the
$l=0$ $\phi$-modes in these STTs. 
The frequencies of the $\phi$-modes 
show again very little dependence on the neutron star mass
as seen already for the $l=2$ $\phi$-modes.
In contrast to the pressure-led modes
their damping times are on the order of ms.

The next steps will include the study of the
$l=1$ dipole modes, and the inclusion of rotation,
at least at a perturbative level. 
Moreover, a whole set of realistic equations of state
will be employed with the aim 
to derive \textit{universal relations}
for the polar modes, analogously to the axial case
\citep{Blazquez-Salcedo:2018qyy}.
Also other closely-related theories as, for example, 
STTs with spontaneous scalarization
will be studied.
 
 Numerical simulations show that 
 the ringdown after a merger is dominated by three modes
 \citep{Bauswein:2015yca,Bernuzzi:2015rla}
 that are interpreted as containing the fundamental f-mode
 and a mixture of the f-mode and a quasiradial F-mode of the
remnant star \citep{Takami:2014tva}.
 The mixing occurs because of the decrease in symmetry 
 when going from spherical to axial symmetry. 
 
 Whereas merger simulations have been mostly performed 
 in General Relativity so far, recently the merger of
 neutron stars has also been considered in STTs
 \citep{Sagunski:2017nzb}.
 Making use of an effective model to estimate the merger of 
 neutron stars in $R^2$ gravity, two modes have been found
 to dominate the ringdown \citep{Sagunski:2017nzb}.  On the other hand, the analysis of the merger of neutron stars in chameleon theory has shown that the $l=1$ dipole modes tend to be suppressed by the screening mechanism \citep{Bezares:2021dma}.
 It will be interesting to perform full merger simulations 
 in STTs to extract the effects of the scalar field
 on the ringdown spectrum and its dependence on the scalar
 field mass.

\section*{Acknowledgments}
We would like to thank Daniela D. Doneva, Burkhard Kleihaus, 
Zahra A.~Motahar and Stoytcho S. Yazadjiev.
We gratefully acknowledge support by the
DFG Research Training Group 1620 {\sl Models of Gravity}, 
the DFG project BL 1553,
and the COST Actions CA15117 {\sl CANTATA} 
and CA16104 {\sl GWverse}. 
JLBS would like to acknowledge support from FCT project
PTDC/FIS-AST/3041/2020.

\section*{Conflict of Interest Statement}

The authors declare that the research was conducted in the absence of any commercial or financial relationships that could be construed as a potential conflict of interest.

\section*{Author Contributions}

All authors have contributed substantially to this paper and
agree to be accountable for the content of the work. 




\section*{Data Availability Statement}
The datasets 
generated and analyzed can be obtained from the authors upon request.

\bibliographystyle{frontiersinSCNS_ENG_HUMS} 
\bibliography{test}

\begin{thebibliography}{108}
\providecommand{\natexlab}[1]{#1}
\expandafter\ifx\csname urlstyle\endcsname\relax
  \providecommand{\doi}[1]{doi:\discretionary{}{}{}#1}\else
  \providecommand{\doi}{doi:\discretionary{}{}{}\begingroup
  \urlstyle{rm}\Url}\fi
\providecommand{\selectlanguage}[1]{\relax}
\providecommand{\bibAnnoteFile}[1]{%
  \IfFileExists{#1}{\begin{quotation}\noindent\textsc{Key:} #1\\
  \textsc{Annotation:}\ \input{#1}\end{quotation}}{}}
\providecommand{\bibAnnote}[2]{%
  \begin{quotation}\noindent\textsc{Key:} #1\\
  \textsc{Annotation:}\ #2\end{quotation}}

\bibitem[{Abbott et~al.(2016{\natexlab{a}})}]{LIGOScientific:2016sjg}
Abbott, B.~P. et~al. (2016{\natexlab{a}}).
\newblock {GW151226: Observation of Gravitational Waves from a 22-Solar-Mass
  Binary Black Hole Coalescence}.
\newblock \emph{Phys. Rev. Lett.} 116, 241103.
\newblock \doi{10.1103/PhysRevLett.116.241103}
\bibAnnoteFile{LIGOScientific:2016sjg}

\bibitem[{Abbott et~al.(2016{\natexlab{b}})}]{LIGOScientific:2016aoc}
Abbott, B.~P. et~al. (2016{\natexlab{b}}).
\newblock {Observation of Gravitational Waves from a Binary Black Hole Merger}.
\newblock \emph{Phys. Rev. Lett.} 116, 061102.
\newblock \doi{10.1103/PhysRevLett.116.061102}
\bibAnnoteFile{LIGOScientific:2016aoc}

\bibitem[{Abbott et~al.(2017{\natexlab{a}})}]{LIGOScientific:2017bnn}
Abbott, B.~P. et~al. (2017{\natexlab{a}}).
\newblock {GW170104: Observation of a 50-Solar-Mass Binary Black Hole
  Coalescence at Redshift 0.2}.
\newblock \emph{Phys. Rev. Lett.} 118, 221101.
\newblock \doi{10.1103/PhysRevLett.118.221101}.
\newblock [Erratum: Phys.Rev.Lett. 121, 129901 (2018)]
\bibAnnoteFile{LIGOScientific:2017bnn}

\bibitem[{Abbott et~al.(2017{\natexlab{b}})}]{LIGOScientific:2017ycc}
Abbott, B.~P. et~al. (2017{\natexlab{b}}).
\newblock {GW170814: A Three-Detector Observation of Gravitational Waves from a
  Binary Black Hole Coalescence}.
\newblock \emph{Phys. Rev. Lett.} 119, 141101.
\newblock \doi{10.1103/PhysRevLett.119.141101}
\bibAnnoteFile{LIGOScientific:2017ycc}

\bibitem[{Abbott et~al.(2017{\natexlab{c}})}]{TheLIGOScientific:2017qsa}
Abbott, B.~P. et~al. (2017{\natexlab{c}}).
\newblock {GW170817: Observation of Gravitational Waves from a Binary Neutron
  Star Inspiral}.
\newblock \emph{Phys. Rev. Lett.} 119, 161101.
\newblock \doi{10.1103/PhysRevLett.119.161101}
\bibAnnoteFile{TheLIGOScientific:2017qsa}

\bibitem[{Abbott et~al.(2017{\natexlab{d}})}]{GBM:2017lvd}
Abbott, B.~P. et~al. (2017{\natexlab{d}}).
\newblock {Multi-messenger Observations of a Binary Neutron Star Merger}.
\newblock \emph{Astrophys. J.} 848, L12.
\newblock \doi{10.3847/2041-8213/aa91c9}
\bibAnnoteFile{GBM:2017lvd}

\bibitem[{Abbott et~al.(2018)}]{LIGOScientific:2017zlf}
Abbott, B.~P. et~al. (2018).
\newblock {GW170817: Implications for the Stochastic Gravitational-Wave
  Background from Compact Binary Coalescences}.
\newblock \emph{Phys. Rev. Lett.} 120, 091101.
\newblock \doi{10.1103/PhysRevLett.120.091101}
\bibAnnoteFile{LIGOScientific:2017zlf}

\bibitem[{Abbott et~al.(2019{\natexlab{a}})}]{LIGOScientific:2018mvr}
Abbott, B.~P. et~al. (2019{\natexlab{a}}).
\newblock {GWTC-1: A Gravitational-Wave Transient Catalog of Compact Binary
  Mergers Observed by LIGO and Virgo during the First and Second Observing
  Runs}.
\newblock \emph{Phys. Rev. X} 9, 031040.
\newblock \doi{10.1103/PhysRevX.9.031040}
\bibAnnoteFile{LIGOScientific:2018mvr}

\bibitem[{Abbott et~al.(2019{\natexlab{b}})}]{Abbott:2018wiz}
Abbott, B.~P. et~al. (2019{\natexlab{b}}).
\newblock {Properties of the binary neutron star merger GW170817}.
\newblock \emph{Phys. Rev.} X9, 011001.
\newblock \doi{10.1103/PhysRevX.9.011001}
\bibAnnoteFile{Abbott:2018wiz}

\bibitem[{Abbott et~al.(2020)}]{LIGOScientific:2020aai}
Abbott, B.~P. et~al. (2020).
\newblock {GW190425: Observation of a Compact Binary Coalescence with Total
  Mass $\sim 3.4 M_{\odot}$}.
\newblock \emph{Astrophys. J. Lett.} 892, L3.
\newblock \doi{10.3847/2041-8213/ab75f5}
\bibAnnoteFile{LIGOScientific:2020aai}

\bibitem[{Altaha~Motahar et~al.(2019)Altaha~Motahar, Bl\'azquez-Salcedo,
  Doneva, Kunz, and Yazadjiev}]{AltahaMotahar:2019ekm}
Altaha~Motahar, Z., Bl\'azquez-Salcedo, J.~L., Doneva, D.~D., Kunz, J., and
  Yazadjiev, S.~S. (2019).
\newblock {Axial quasinormal modes of scalarized neutron stars with massive
  self-interacting scalar field}.
\newblock \emph{Phys. Rev. D} 99, 104006.
\newblock \doi{10.1103/PhysRevD.99.104006}
\bibAnnoteFile{AltahaMotahar:2019ekm}

\bibitem[{Altaha~Motahar et~al.(2018)Altaha~Motahar, Blázquez-Salcedo,
  Kleihaus, and Kunz}]{AltahaMotahar:2018djk}
Altaha~Motahar, Z., Blázquez-Salcedo, J.~L., Kleihaus, B., and Kunz, J.
  (2018).
\newblock {Axial quasinormal modes of scalarized neutron stars with realistic
  equations of state}.
\newblock \emph{Phys. Rev.} D98, 044032.
\newblock \doi{10.1103/PhysRevD.98.044032}
\bibAnnoteFile{AltahaMotahar:2018djk}

\bibitem[{Andersson and Kokkotas(1996)}]{Andersson:1996pn}
Andersson, N. and Kokkotas, K.~D. (1996).
\newblock {Gravitational waves and pulsating stars: What can we learn from
  future observations?}
\newblock \emph{Phys. Rev. Lett.} 77, 4134--4137.
\newblock \doi{10.1103/PhysRevLett.77.4134}
\bibAnnoteFile{Andersson:1996pn}

\bibitem[{Andersson and Kokkotas(1998)}]{Andersson:1997rn}
Andersson, N. and Kokkotas, K.~D. (1998).
\newblock {Towards gravitational wave asteroseismology}.
\newblock \emph{Mon. Not. Roy. Astron. Soc.} 299, 1059--1068.
\newblock \doi{10.1046/j.1365-8711.1998.01840.x}
\bibAnnoteFile{Andersson:1997rn}

\bibitem[{Antoniadis et~al.(2013)}]{Antoniadis:2013pzd}
Antoniadis, J. et~al. (2013).
\newblock {A Massive Pulsar in a Compact Relativistic Binary}.
\newblock \emph{Science} 340, 6131.
\newblock \doi{10.1126/science.1233232}
\bibAnnoteFile{Antoniadis:2013pzd}

\bibitem[{Ascher et~al.(1979)Ascher, Christiansen, and
  Russell}]{Ascher:1979iha}
Ascher, U., Christiansen, J., and Russell, R.~D. (1979).
\newblock {A Collocation Solver for Mixed Order Systems of Boundary Value
  Problems}.
\newblock \emph{Math. Comput.} 33, 659--679.
\newblock \doi{10.1090/S0025-5718-1979-0521281-7}
\bibAnnoteFile{Ascher:1979iha}

\bibitem[{Astashenok et~al.(2017)Astashenok, Odintsov, and de~la
  Cruz-Dombriz}]{Astashenok:2017dpo}
Astashenok, A.~V., Odintsov, S.~D., and de~la Cruz-Dombriz, A. (2017).
\newblock {The realistic models of relativistic stars in $f(R) = R + \alpha
  R^2$ gravity}.
\newblock \emph{Class. Quant. Grav.} 34, 205008.
\newblock \doi{10.1088/1361-6382/aa8971}
\bibAnnoteFile{Astashenok:2017dpo}

\bibitem[{Barack et~al.(2019)}]{Barack:2018yly}
Barack, L. et~al. (2019).
\newblock {Black holes, gravitational waves and fundamental physics: a
  roadmap}.
\newblock \emph{Class. Quant. Grav.} 36, 143001.
\newblock \doi{10.1088/1361-6382/ab0587}
\bibAnnoteFile{Barack:2018yly}

\bibitem[{{Bardeen} et~al.(1966){Bardeen}, {Thorne}, and
  {Meltzer}}]{1966ApJ...145..505B}
{Bardeen}, J.~M., {Thorne}, K.~S., and {Meltzer}, D.~W. (1966).
\newblock {A Catalogue of Methods for Studying the Normal Modes of Radial
  Pulsation of General-Relativistic Stellar Models}.
\newblock \emph{Astrophysical Journal} 145, 505.
\newblock \doi{10.1086/148791}
\bibAnnoteFile{1966ApJ...145..505B}

\bibitem[{Bauswein and Stergioulas(2015)}]{Bauswein:2015yca}
Bauswein, A. and Stergioulas, N. (2015).
\newblock {Unified picture of the post-merger dynamics and gravitational wave
  emission in neutron star mergers}.
\newblock \emph{Phys. Rev.} D91, 124056.
\newblock \doi{10.1103/PhysRevD.91.124056}
\bibAnnoteFile{Bauswein:2015yca}

\bibitem[{Benhar et~al.(2004)Benhar, Ferrari, and Gualtieri}]{Benhar:2004xg}
Benhar, O., Ferrari, V., and Gualtieri, L. (2004).
\newblock {Gravitational wave asteroseismology revisited}.
\newblock \emph{Phys. Rev.} D70, 124015.
\newblock \doi{10.1103/PhysRevD.70.124015}
\bibAnnoteFile{Benhar:2004xg}

\bibitem[{Bernuzzi et~al.(2015)Bernuzzi, Dietrich, and
  Nagar}]{Bernuzzi:2015rla}
Bernuzzi, S., Dietrich, T., and Nagar, A. (2015).
\newblock {Modeling the complete gravitational wave spectrum of neutron star
  mergers}.
\newblock \emph{Phys. Rev. Lett.} 115, 091101.
\newblock \doi{10.1103/PhysRevLett.115.091101}
\bibAnnoteFile{Bernuzzi:2015rla}

\bibitem[{Berti et~al.(2009)Berti, Cardoso, and Starinets}]{Berti:2009kk}
Berti, E., Cardoso, V., and Starinets, A.~O. (2009).
\newblock {Quasinormal modes of black holes and black branes}.
\newblock \emph{Class. Quant. Grav.} 26, 163001.
\newblock \doi{10.1088/0264-9381/26/16/163001}
\bibAnnoteFile{Berti:2009kk}

\bibitem[{Berti et~al.(2018{\natexlab{a}})Berti, Yagi, Yang, and
  Yunes}]{Berti:2018vdi}
Berti, E., Yagi, K., Yang, H., and Yunes, N. (2018{\natexlab{a}}).
\newblock {Extreme Gravity Tests with Gravitational Waves from Compact Binary
  Coalescences: (II) Ringdown}.
\newblock \emph{Gen. Rel. Grav.} 50, 49.
\newblock \doi{10.1007/s10714-018-2372-6}
\bibAnnoteFile{Berti:2018vdi}

\bibitem[{Berti et~al.(2018{\natexlab{b}})Berti, Yagi, and
  Yunes}]{Berti:2018cxi}
Berti, E., Yagi, K., and Yunes, N. (2018{\natexlab{b}}).
\newblock {Extreme Gravity Tests with Gravitational Waves from Compact Binary
  Coalescences: (I) Inspiral-Merger}.
\newblock \emph{Gen. Rel. Grav.} 50, 46.
\newblock \doi{10.1007/s10714-018-2362-8}
\bibAnnoteFile{Berti:2018cxi}

\bibitem[{Berti et~al.(2015)}]{Berti:2015itd}
Berti, E. et~al. (2015).
\newblock {Testing General Relativity with Present and Future Astrophysical
  Observations}.
\newblock \emph{Class. Quant. Grav.} 32, 243001.
\newblock \doi{10.1088/0264-9381/32/24/243001}
\bibAnnoteFile{Berti:2015itd}

\bibitem[{Bezares et~al.(2021)Bezares, Aguilera-Miret, ter Haar, Crisostomi,
  Palenzuela, and Barausse}]{Bezares:2021dma}
Bezares, M., Aguilera-Miret, R., ter Haar, L., Crisostomi, M., Palenzuela, C.,
  and Barausse, E. (2021).
\newblock {No evidence of kinetic screening in merging binary neutron stars}
\bibAnnoteFile{Bezares:2021dma}

\bibitem[{Bhagwat et~al.(2020)Bhagwat, Cabero, Capano, Krishnan, and
  Brown}]{Bhagwat:2019bwv}
Bhagwat, S., Cabero, M., Capano, C.~D., Krishnan, B., and Brown, D.~A. (2020).
\newblock {Detectability of the subdominant mode in a binary black hole
  ringdown}.
\newblock \emph{Phys. Rev. D} 102, 024023.
\newblock \doi{10.1103/PhysRevD.102.024023}
\bibAnnoteFile{Bhagwat:2019bwv}

\bibitem[{Bhattacharya and Majhi(2017)}]{Bhattacharya:2017pqc}
Bhattacharya, K. and Majhi, B.~R. (2017).
\newblock {Fresh look at the scalar-tensor theory of gravity in Jordan and
  Einstein frames from undiscussed standpoints}.
\newblock \emph{Phys. Rev.} D95, 064026.
\newblock \doi{10.1103/PhysRevD.95.064026}
\bibAnnoteFile{Bhattacharya:2017pqc}

\bibitem[{Bl\'azquez-Salcedo et~al.(2019)Bl\'azquez-Salcedo, Altaha~Motahar,
  Doneva, Khoo, Kunz, Mojica et~al.}]{Blazquez-Salcedo:2018pxo}
Bl\'azquez-Salcedo, J.~L., Altaha~Motahar, Z., Doneva, D.~D., Khoo, F.~S.,
  Kunz, J., Mojica, S., et~al. (2019).
\newblock {Quasinormal modes of compact objects in alternative theories of
  gravity}.
\newblock \emph{Eur. Phys. J. Plus} 134, 46.
\newblock \doi{10.1140/epjp/i2019-12392-9}
\bibAnnoteFile{Blazquez-Salcedo:2018pxo}

\bibitem[{Bl\'azquez-Salcedo et~al.(2020)Bl\'azquez-Salcedo, Scen~Khoo, and
  Kunz}]{Blazquez-Salcedo:2020ibb}
Bl\'azquez-Salcedo, J.~L., Scen~Khoo, F., and Kunz, J. (2020).
\newblock {Ultra-long-lived quasi-normal modes of neutron stars in massive
  scalar-tensor gravity}.
\newblock \emph{EPL} 130, 50002.
\newblock \doi{10.1209/0295-5075/130/50002}
\bibAnnoteFile{Blazquez-Salcedo:2020ibb}

\bibitem[{Blázquez-Salcedo et~al.(2018)Blázquez-Salcedo, Doneva, Kunz,
  Staykov, and Yazadjiev}]{Blazquez-Salcedo:2018qyy}
Blázquez-Salcedo, J.~L., Doneva, D.~D., Kunz, J., Staykov, K.~V., and
  Yazadjiev, S.~S. (2018).
\newblock {Axial quasinormal modes of neutron stars in $R^2$ gravity}.
\newblock \emph{Phys. Rev.} D98, 104047.
\newblock \doi{10.1103/PhysRevD.98.104047}
\bibAnnoteFile{Blazquez-Salcedo:2018qyy}

\bibitem[{Blázquez-Salcedo and Eickhoff(2018)}]{Blazquez-Salcedo:2018tyn}
Blázquez-Salcedo, J.~L. and Eickhoff, K. (2018).
\newblock {Axial quasinormal modes of static neutron stars in the nonminimal
  derivative coupling sector of Horndeski gravity: spectrum and universal
  relations for realistic equations of state}.
\newblock \emph{Phys. Rev.} D97, 104002.
\newblock \doi{10.1103/PhysRevD.97.104002}
\bibAnnoteFile{Blazquez-Salcedo:2018tyn}

\bibitem[{Blázquez-Salcedo et~al.(2016{\natexlab{a}})Blázquez-Salcedo,
  González-Romero, Kunz, Mojica, and
  Navarro-Lérida}]{Blazquez-Salcedo:2015ets}
Blázquez-Salcedo, J.~L., González-Romero, L.~M., Kunz, J., Mojica, S., and
  Navarro-Lérida, F. (2016{\natexlab{a}}).
\newblock {Axial quasinormal modes of Einstein-Gauss-Bonnet-dilaton neutron
  stars}.
\newblock \emph{Phys. Rev.} D93, 024052.
\newblock \doi{10.1103/PhysRevD.93.024052}
\bibAnnoteFile{Blazquez-Salcedo:2015ets}

\bibitem[{Blázquez-Salcedo et~al.(2013)Blázquez-Salcedo, González-Romero,
  and Navarro-Lérida}]{BlazquezSalcedo:2012pd}
Blázquez-Salcedo, J.~L., González-Romero, L.~M., and Navarro-Lérida, F.
  (2013).
\newblock {Phenomenological relations for axial quasinormal modes of neutron
  stars with realistic equations of state}.
\newblock \emph{Phys. Rev.} D87, 104042.
\newblock \doi{10.1103/PhysRevD.87.104042}
\bibAnnoteFile{BlazquezSalcedo:2012pd}

\bibitem[{Blázquez-Salcedo et~al.(2014)Blázquez-Salcedo, González-Romero,
  and Navarro-Lérida}]{Blazquez-Salcedo:2013jka}
Blázquez-Salcedo, J.~L., González-Romero, L.~M., and Navarro-Lérida, F.
  (2014).
\newblock {Polar quasi-normal modes of neutron stars with equations of state
  satisfying the $2 M_{\odot}$ constraint}.
\newblock \emph{Phys. Rev.} D89, 044006.
\newblock \doi{10.1103/PhysRevD.89.044006}
\bibAnnoteFile{Blazquez-Salcedo:2013jka}

\bibitem[{Blázquez-Salcedo et~al.(2017)Blázquez-Salcedo, Khoo, and
  Kunz}]{Blazquez-Salcedo:2017txk}
Blázquez-Salcedo, J.~L., Khoo, F.~S., and Kunz, J. (2017).
\newblock {Quasinormal modes of Einstein-Gauss-Bonnet-dilaton black holes}.
\newblock \emph{Phys. Rev.} D96, 064008.
\newblock \doi{10.1103/PhysRevD.96.064008}
\bibAnnoteFile{Blazquez-Salcedo:2017txk}

\bibitem[{Blázquez-Salcedo et~al.(2016{\natexlab{b}})Blázquez-Salcedo,
  Macedo, Cardoso, Ferrari, Gualtieri, Khoo et~al.}]{Blazquez-Salcedo:2016enn}
Blázquez-Salcedo, J.~L., Macedo, C. F.~B., Cardoso, V., Ferrari, V.,
  Gualtieri, L., Khoo, F.~S., et~al. (2016{\natexlab{b}}).
\newblock {Perturbed black holes in Einstein-dilaton-Gauss-Bonnet gravity:
  Stability, ringdown, and gravitational-wave emission}.
\newblock \emph{Phys. Rev.} D94, 104024.
\newblock \doi{10.1103/PhysRevD.94.104024}
\bibAnnoteFile{Blazquez-Salcedo:2016enn}

\bibitem[{Brito et~al.(2017)Brito, Ghosh, Barausse, Berti, Cardoso, Dvorkin
  et~al.}]{Brito:2017zvb}
Brito, R., Ghosh, S., Barausse, E., Berti, E., Cardoso, V., Dvorkin, I., et~al.
  (2017).
\newblock {Gravitational wave searches for ultralight bosons with LIGO and
  LISA}.
\newblock \emph{Phys. Rev.} D96, 064050.
\newblock \doi{10.1103/PhysRevD.96.064050}
\bibAnnoteFile{Brito:2017zvb}

\bibitem[{{Campolattaro} and {Thorne}(1970)}]{1970ApJ...159..847C}
{Campolattaro}, A. and {Thorne}, K.~S. (1970).
\newblock {Nonradial Pulsation of General-Relativistic Stellar Models. V.
  Analytic Analysis for L = 1}.
\newblock \emph{Astrophys. J.} 159, 847.
\newblock \doi{10.1086/150362}
\bibAnnoteFile{1970ApJ...159..847C}

\bibitem[{Capano et~al.(2021)Capano, Cabero, Westerweck, Abedi, Kastha, Nitz
  et~al.}]{Capano:2021etf}
Capano, C.~D., Cabero, M., Westerweck, J., Abedi, J., Kastha, S., Nitz, A.~H.,
  et~al. (2021).
\newblock {Observation of a multimode quasi-normal spectrum from a perturbed
  black hole}
\bibAnnoteFile{Capano:2021etf}

\bibitem[{Capozziello and De~Laurentis(2011)}]{Capozziello:2011et}
Capozziello, S. and De~Laurentis, M. (2011).
\newblock {Extended Theories of Gravity}.
\newblock \emph{Phys. Rept.} 509, 167--321.
\newblock \doi{10.1016/j.physrep.2011.09.003}
\bibAnnoteFile{Capozziello:2011et}

\bibitem[{Chandrasekhar(1964{\natexlab{a}})}]{Chandrasekhar:1964zza}
Chandrasekhar, S. (1964{\natexlab{a}}).
\newblock {Dynamical Instability of Gaseous Masses Approaching the
  Schwarzschild Limit in General Relativity}.
\newblock \emph{Phys. Rev. Lett.} 12, 114--116.
\newblock \doi{10.1103/PhysRevLett.12.114}
\bibAnnoteFile{Chandrasekhar:1964zza}

\bibitem[{Chandrasekhar(1964{\natexlab{b}})}]{Chandrasekhar:1964zz}
Chandrasekhar, S. (1964{\natexlab{b}}).
\newblock {The Dynamical Instability of Gaseous Masses Approaching the
  Schwarzschild Limit in General Relativity}.
\newblock \emph{Astrophys. J.} 140, 417--433.
\newblock \doi{10.1086/147938}.
\newblock [Erratum: Astrophys. J.140,1342(1964)]
\bibAnnoteFile{Chandrasekhar:1964zz}

\bibitem[{Chandrasekhar and Ferrari(1991)}]{doi:10.1098/rspa.1991.0016}
Chandrasekhar, S. and Ferrari, V. (1991).
\newblock On the non-radial oscillations of a star.
\newblock \emph{Proceedings of the Royal Society of London. Series A:
  Mathematical and Physical Sciences} 432, 247--279.
\newblock \doi{10.1098/rspa.1991.0016}
\bibAnnoteFile{doi:10.1098/rspa.1991.0016}

\bibitem[{Chandrasekhar et~al.(1991{\natexlab{a}})Chandrasekhar, Ferrari, and
  Enderby}]{doi:10.1098/rspa.1991.0104}
Chandrasekhar, S., Ferrari, V., and Enderby, J.~E. (1991{\natexlab{a}}).
\newblock On the non-radial oscillations of a star. iii. a reconsideration of
  the axial modes.
\newblock \emph{Proceedings of the Royal Society of London. Series A:
  Mathematical and Physical Sciences} 434, 449--457.
\newblock \doi{10.1098/rspa.1991.0104}
\bibAnnoteFile{doi:10.1098/rspa.1991.0104}

\bibitem[{Chandrasekhar et~al.(1991{\natexlab{b}})Chandrasekhar, Ferrari, and
  Winston}]{doi:10.1098/rspa.1991.0117}
Chandrasekhar, S., Ferrari, V., and Winston, R. (1991{\natexlab{b}}).
\newblock On the non-radial oscillations of a star - ii. further
  amplifications.
\newblock \emph{Proceedings of the Royal Society of London. Series A:
  Mathematical and Physical Sciences} 434, 635--641.
\newblock \doi{10.1098/rspa.1991.0117}
\bibAnnoteFile{doi:10.1098/rspa.1991.0117}

\bibitem[{{Chanmugam}(1977)}]{1977ApJ...217..799C}
{Chanmugam}, G. (1977).
\newblock {Radial oscillations of zero-temperature white dwarfs and neutron
  stars below nuclear densities.}
\newblock \emph{Astrophysical Journal} 217, 799--808.
\newblock \doi{10.1086/155627}
\bibAnnoteFile{1977ApJ...217..799C}

\bibitem[{Coulter et~al.(2017)}]{Coulter:2017wya}
Coulter, D.~A. et~al. (2017).
\newblock {Swope Supernova Survey 2017a (SSS17a), the Optical Counterpart to a
  Gravitational Wave Source}.
\newblock \emph{Science} \doi{10.1126/science.aap9811}.
\newblock [Science358,1556(2017)]
\bibAnnoteFile{Coulter:2017wya}

\bibitem[{{Datta} et~al.(1998){Datta}, {Hasan}, {Sahu}, and
  {Prasanna}}]{1998IJMPD...7...49D}
{Datta}, B., {Hasan}, S.~S., {Sahu}, P.~K., and {Prasanna}, A.~R. (1998).
\newblock {Radial Modes of Rotating Neutron Stars in the Chandrasekhar-Friedman
  Formalism}.
\newblock \emph{International Journal of Modern Physics D} 7, 49--59.
\newblock \doi{10.1142/S021827189800005X}
\bibAnnoteFile{1998IJMPD...7...49D}

\bibitem[{De~Felice and Tsujikawa(2010)}]{DeFelice:2010aj}
De~Felice, A. and Tsujikawa, S. (2010).
\newblock {f(R) theories}.
\newblock \emph{Living Rev. Rel.} 13, 3.
\newblock \doi{10.12942/lrr-2010-3}
\bibAnnoteFile{DeFelice:2010aj}

\bibitem[{Detweiler and Lindblom(1985)}]{Detweiler:1985zz}
Detweiler, S.~L. and Lindblom, L. (1985).
\newblock {On the nonradial pulsations of general relativistic stellar models}.
\newblock \emph{Astrophys. J.} 292, 12--15.
\newblock \doi{10.1086/163127}
\bibAnnoteFile{Detweiler:1985zz}

\bibitem[{Dima et~al.(2021)Dima, Bezares, and Barausse}]{Dima:2021pwx}
Dima, A., Bezares, M., and Barausse, E. (2021).
\newblock {Dynamical Chameleon Neutron Stars: stability, radial oscillations
  and scalar radiation in spherical symmetry}
\bibAnnoteFile{Dima:2021pwx}

\bibitem[{Doneva and Pappas(2018)}]{Doneva:2017jop}
Doneva, D.~D. and Pappas, G. (2018).
\newblock {Universal Relations and Alternative Gravity Theories}.
\newblock \emph{Astrophys. Space Sci. Libr.} 457, 737--806.
\newblock \doi{10.1007/978-3-319-97616-7_13}
\bibAnnoteFile{Doneva:2017jop}

\bibitem[{Douchin and Haensel(2001)}]{Douchin:2001sv}
Douchin, F. and Haensel, P. (2001).
\newblock {A unified equation of state of dense matter and neutron star
  structure}.
\newblock \emph{Astron. Astrophys.} 380, 151.
\newblock \doi{10.1051/0004-6361:20011402}
\bibAnnoteFile{Douchin:2001sv}

\bibitem[{Faraoni(2009)}]{Faraoni:2009km}
Faraoni, V. (2009).
\newblock {Scalar field mass in generalized gravity}.
\newblock \emph{Class. Quant. Grav.} 26, 145014.
\newblock \doi{10.1088/0264-9381/26/14/145014}
\bibAnnoteFile{Faraoni:2009km}

\bibitem[{Faraoni and Capozziello(2011)}]{Faraoni:2010pgm}
Faraoni, V. and Capozziello, S. (2011).
\newblock \emph{{Beyond Einstein Gravity}: {A Survey of Gravitational Theories
  for Cosmology and Astrophysics}} (Dordrecht: Springer).
\newblock \doi{10.1007/978-94-007-0165-6}
\bibAnnoteFile{Faraoni:2010pgm}

\bibitem[{Faraoni and Gunzig(1999)}]{Faraoni:1999hp}
Faraoni, V. and Gunzig, E. (1999).
\newblock {Einstein frame or Jordan frame?}
\newblock \emph{Int. J. Theor. Phys.} 38, 217--225.
\newblock \doi{10.1023/A:1026645510351}
\bibAnnoteFile{Faraoni:1999hp}

\bibitem[{Fernandez-Jambrina and
  Gonzalez-Romero(2003)}]{FernandezJambrina:2003mv}
Fernandez-Jambrina, L. and Gonzalez-Romero, L.~M. (eds.) (2003).
\newblock \emph{{Current trends in relativistic astrophysics: Theoretical,
  numerical, observational. Proceedings, 24th Meeting, ERE 2001, Madrid, Spain,
  September 18-21, 2001}}, vol. 617.
\newblock \doi{10.1007/3-540-36973-2}
\bibAnnoteFile{FernandezJambrina:2003mv}

\bibitem[{Giesler et~al.(2019)Giesler, Isi, Scheel, and
  Teukolsky}]{Giesler:2019uxc}
Giesler, M., Isi, M., Scheel, M.~A., and Teukolsky, S. (2019).
\newblock {Black Hole Ringdown: The Importance of Overtones}.
\newblock \emph{Phys. Rev. X} 9, 041060.
\newblock \doi{10.1103/PhysRevX.9.041060}
\bibAnnoteFile{Giesler:2019uxc}

\bibitem[{{Glass} and {Lindblom}(1983)}]{1983ApJS...53...93G}
{Glass}, E.~N. and {Lindblom}, L. (1983).
\newblock {The Radial Oscillations of Neutron Stars}.
\newblock \emph{Astrophysical Journal, Supplement} 53, 93.
\newblock \doi{10.1086/190885}
\bibAnnoteFile{1983ApJS...53...93G}

\bibitem[{Haensel et~al.(2007)Haensel, Potekhin, and Yakovlev}]{Haensel:2007yy}
Haensel, P., Potekhin, A.~Y., and Yakovlev, D.~G. (2007).
\newblock {Neutron stars 1: Equation of state and structure}.
\newblock \emph{Astrophys. Space Sci. Libr.} 326, pp.1--619.
\newblock \doi{10.1007/978-0-387-47301-7}
\bibAnnoteFile{Haensel:2007yy}

\bibitem[{Ipser and Price(1991)}]{PhysRevD.43.1768}
Ipser, J.~R. and Price, R.~H. (1991).
\newblock Nonradial pulsations of stellar models in general relativity.
\newblock \emph{Phys. Rev. D} 43, 1768--1773.
\newblock \doi{10.1103/PhysRevD.43.1768}
\bibAnnoteFile{PhysRevD.43.1768}

\bibitem[{Jim\'enez~Forteza et~al.(2020)Jim\'enez~Forteza, Bhagwat, Pani, and
  Ferrari}]{JimenezForteza:2020cve}
Jim\'enez~Forteza, X., Bhagwat, S., Pani, P., and Ferrari, V. (2020).
\newblock {Spectroscopy of binary black hole ringdown using overtones and
  angular modes}.
\newblock \emph{Phys. Rev. D} 102, 044053.
\newblock \doi{10.1103/PhysRevD.102.044053}
\bibAnnoteFile{JimenezForteza:2020cve}

\bibitem[{Kojima(1992)}]{Kojima:1992ie}
Kojima, Y. (1992).
\newblock {Equations governing the nonradial oscillations of a slowly rotating
  relativistic star}.
\newblock \emph{Phys. Rev.} D46, 4289--4303.
\newblock \doi{10.1103/PhysRevD.46.4289}
\bibAnnoteFile{Kojima:1992ie}

\bibitem[{Kokkotas et~al.(2001)Kokkotas, Apostolatos, and
  Andersson}]{Kokkotas:1999mn}
Kokkotas, K.~D., Apostolatos, T.~A., and Andersson, N. (2001).
\newblock {The Inverse problem for pulsating neutron stars: A 'Fingerprint
  analysis' for the supranuclear equation of state}.
\newblock \emph{Mon. Not. Roy. Astron. Soc.} 320, 307--315.
\newblock \doi{10.1046/j.1365-8711.2001.03945.x}
\bibAnnoteFile{Kokkotas:1999mn}

\bibitem[{Kokkotas and Schmidt(1999)}]{Kokkotas:1999bd}
Kokkotas, K.~D. and Schmidt, B.~G. (1999).
\newblock {Quasinormal modes of stars and black holes}.
\newblock \emph{Living Rev. Rel.} 2, 2.
\newblock \doi{10.12942/lrr-1999-2}
\bibAnnoteFile{Kokkotas:1999bd}

\bibitem[{Kokkotas and Schutz(1986)}]{Kokkotas:1986gd}
Kokkotas, K.~D. and Schutz, B.~F. (1986).
\newblock {Normal Modes of a Model Radiating System}.
\newblock \emph{Gen. Rel. Grav.} 18, 913
\bibAnnoteFile{Kokkotas:1986gd}

\bibitem[{Kokkotas and Schutz(1992)}]{Kokkotas:2003mh}
Kokkotas, K.~D. and Schutz, B.~F. (1992).
\newblock {W-modes: A New family of normal modes of pulsating relativistic
  stars}.
\newblock \emph{Mon. Not. Roy. Astron. Soc.} 255, 119
\bibAnnoteFile{Kokkotas:2003mh}

\bibitem[{Konoplya and Zhidenko(2011)}]{Konoplya:2011qq}
Konoplya, R.~A. and Zhidenko, A. (2011).
\newblock {Quasinormal modes of black holes: From astrophysics to string
  theory}.
\newblock \emph{Rev. Mod. Phys.} 83, 793--836.
\newblock \doi{10.1103/RevModPhys.83.793}
\bibAnnoteFile{Konoplya:2011qq}

\bibitem[{Kr\"uger and Doneva(2021)}]{Kruger:2021yay}
Kr\"uger, C.~J. and Doneva, D.~D. (2021).
\newblock {Oscillation dynamics of scalarized neutron stars}.
\newblock \emph{Phys. Rev. D} 103, 124034.
\newblock \doi{10.1103/PhysRevD.103.124034}
\bibAnnoteFile{Kruger:2021yay}

\bibitem[{Lattimer and Steiner(2014)}]{Lattimer:2013hma}
Lattimer, J.~M. and Steiner, A.~W. (2014).
\newblock {Neutron Star Masses and Radii from Quiescent Low-Mass X-ray
  Binaries}.
\newblock \emph{Astrophys. J.} 784, 123.
\newblock \doi{10.1088/0004-637X/784/2/123}
\bibAnnoteFile{Lattimer:2013hma}

\bibitem[{Lindblom and Detweiler(1983)}]{Lindblom:1983ps}
Lindblom, L. and Detweiler, S.~L. (1983).
\newblock {The quadrupole oscillations of neutron stars}.
\newblock \emph{Astrophys. J. Suppl.} 53, 73--92.
\newblock \doi{10.1086/190884}
\bibAnnoteFile{Lindblom:1983ps}

\bibitem[{{Meltzer} and {Thorne}(1966)}]{1966ApJ...145..514M}
{Meltzer}, D.~W. and {Thorne}, K.~S. (1966).
\newblock {Normal Modes of Radial Pulsation of Stars at the End Point of
  Thermonuclear Evolution}.
\newblock \emph{Astrophysical Journal} 145, 514.
\newblock \doi{10.1086/148792}
\bibAnnoteFile{1966ApJ...145..514M}

\bibitem[{Mena-Fernández and González-Romero(2019)}]{Mena-Fernandez:2019irg}
Mena-Fernández, J. and González-Romero, L.~M. (2019).
\newblock {Reconstruction of the neutron star equation of state from
  w-quasinormal modes spectra with a piecewise polytropic meshing and
  refinement method}.
\newblock \emph{arXiv:1901.10851}
\bibAnnoteFile{Mena-Fernandez:2019irg}

\bibitem[{Mendes and Ortiz(2018)}]{Mendes:2018qwo}
Mendes, R. F.~P. and Ortiz, N. (2018).
\newblock {New class of quasinormal modes of neutron stars in scalar-tensor
  gravity}.
\newblock \emph{Phys. Rev. Lett.} 120, 201104.
\newblock \doi{10.1103/PhysRevLett.120.201104}
\bibAnnoteFile{Mendes:2018qwo}

\bibitem[{Most et~al.(2018)Most, Weih, Rezzolla, and
  Schaffner-Bielich}]{Most:2018hfd}
Most, E.~R., Weih, L.~R., Rezzolla, L., and Schaffner-Bielich, J. (2018).
\newblock {New constraints on radii and tidal deformabilities of neutron stars
  from GW170817}.
\newblock \emph{Phys. Rev. Lett.} 120, 261103.
\newblock \doi{10.1103/PhysRevLett.120.261103}
\bibAnnoteFile{Most:2018hfd}

\bibitem[{Naf and Jetzer(2010)}]{Naf:2010zy}
Naf, J. and Jetzer, P. (2010).
\newblock {On the 1/c Expansion of f(R) Gravity}.
\newblock \emph{Phys. Rev. D} 81, 104003.
\newblock \doi{10.1103/PhysRevD.81.104003}
\bibAnnoteFile{Naf:2010zy}

\bibitem[{Nollert(1999)}]{Nollert:1999ji}
Nollert, H.-P. (1999).
\newblock {TOPICAL REVIEW: Quasinormal modes: the characteristic `sound' of
  black holes and neutron stars}.
\newblock \emph{Class. Quant. Grav.} 16, R159--R216.
\newblock \doi{10.1088/0264-9381/16/12/201}
\bibAnnoteFile{Nollert:1999ji}

\bibitem[{Orellana et~al.(2013)Orellana, Garcia, Teppa~Pannia, and
  Romero}]{Orellana:2013gn}
Orellana, M., Garcia, F., Teppa~Pannia, F.~A., and Romero, G.~E. (2013).
\newblock {Structure of neutron stars in $R$-squared gravity}.
\newblock \emph{Gen. Rel. Grav.} 45, 771--783.
\newblock \doi{10.1007/s10714-013-1501-5}
\bibAnnoteFile{Orellana:2013gn}

\bibitem[{{Price} and {Thorne}(1969)}]{1969ApJ...155..163P}
{Price}, R. and {Thorne}, K.~S. (1969).
\newblock {Non-Radial Pulsation of General-Relativistic Stellar Models. II.
  Properties of the Gravitational Waves}.
\newblock \emph{Astrophys. J.} 155, 163.
\newblock \doi{10.1086/149857}
\bibAnnoteFile{1969ApJ...155..163P}

\bibitem[{Read et~al.(2009)Read, Lackey, Owen, and Friedman}]{Read:2008iy}
Read, J.~S., Lackey, B.~D., Owen, B.~J., and Friedman, J.~L. (2009).
\newblock {Constraints on a phenomenologically parameterized neutron-star
  equation of state}.
\newblock \emph{Phys. Rev.} D79, 124032.
\newblock \doi{10.1103/PhysRevD.79.124032}
\bibAnnoteFile{Read:2008iy}

\bibitem[{{Regge} and {Wheeler}(1957)}]{ReggeW}
{Regge}, T. and {Wheeler}, J.~A. (1957).
\newblock {Stability of a Schwarzschild Singularity}.
\newblock \emph{Phys. Rev.} 108, 1063--1069.
\newblock \doi{10.1103/PhysRev.108.1063}
\bibAnnoteFile{ReggeW}

\bibitem[{Rosca-Mead(2020)}]{Rosca-Mead:2020cyo}
Rosca-Mead, R. (2020).
\newblock \emph{{Gravitational collapse, compact objects and gravitational
  waves in General Relativity and modified gravity}}.
\newblock Ph.D. thesis, University of Cambridge, Department of Applied
  Mathematics and Theoretical Physics, Newnham.
\newblock \doi{10.17863/CAM.53747}
\bibAnnoteFile{Rosca-Mead:2020cyo}

\bibitem[{Rosca-Mead et~al.(2020{\natexlab{a}})Rosca-Mead, Moore, Sperhake,
  Agathos, and Gerosa}]{Rosca-Mead:2020bzt}
Rosca-Mead, R., Moore, C.~J., Sperhake, U., Agathos, M., and Gerosa, D.
  (2020{\natexlab{a}}).
\newblock {Structure of neutron stars in massive scalar-tensor gravity}.
\newblock \emph{Symmetry} 12, 1384.
\newblock \doi{10.3390/sym12091384}
\bibAnnoteFile{Rosca-Mead:2020bzt}

\bibitem[{Rosca-Mead et~al.(2020{\natexlab{b}})Rosca-Mead, Sperhake, Moore,
  Agathos, Gerosa, and Ott}]{Rosca-Mead:2020ehn}
Rosca-Mead, R., Sperhake, U., Moore, C.~J., Agathos, M., Gerosa, D., and Ott,
  C.~D. (2020{\natexlab{b}}).
\newblock {Core collapse in massive scalar-tensor gravity}.
\newblock \emph{Phys. Rev. D} 102, 044010.
\newblock \doi{10.1103/PhysRevD.102.044010}
\bibAnnoteFile{Rosca-Mead:2020ehn}

\bibitem[{Sagunski et~al.(2018)Sagunski, Zhang, Johnson, Lehner, Sakellariadou,
  Liebling et~al.}]{Sagunski:2017nzb}
Sagunski, L., Zhang, J., Johnson, M.~C., Lehner, L., Sakellariadou, M.,
  Liebling, S.~L., et~al. (2018).
\newblock {Neutron star mergers as a probe of modifications of general
  relativity with finite-range scalar forces}.
\newblock \emph{Phys. Rev.} D97, 064016.
\newblock \doi{10.1103/PhysRevD.97.064016}
\bibAnnoteFile{Sagunski:2017nzb}

\bibitem[{Saridakis et~al.(2021)}]{CANTATA:2021ktz}
Saridakis, E.~N. et~al. (2021).
\newblock {Modified Gravity and Cosmology: An Update by the CANTATA Network}
\bibAnnoteFile{CANTATA:2021ktz}

\bibitem[{Sotani and Kokkotas(2004)}]{Sotani:2004rq}
Sotani, H. and Kokkotas, K.~D. (2004).
\newblock {Probing strong-field scalar-tensor gravity with gravitational wave
  asteroseismology}.
\newblock \emph{Phys. Rev.} D70, 084026.
\newblock \doi{10.1103/PhysRevD.70.084026}
\bibAnnoteFile{Sotani:2004rq}

\bibitem[{Sotani and Kokkotas(2005)}]{Sotani:2005qx}
Sotani, H. and Kokkotas, K.~D. (2005).
\newblock {Stellar oscillations in scalar-tensor theory of gravity}.
\newblock \emph{Phys. Rev. D} 71, 124038.
\newblock \doi{10.1103/PhysRevD.71.124038}
\bibAnnoteFile{Sotani:2005qx}

\bibitem[{Sotiriou and Faraoni(2010)}]{Sotiriou:2008rp}
Sotiriou, T.~P. and Faraoni, V. (2010).
\newblock {f(R) Theories Of Gravity}.
\newblock \emph{Rev. Mod. Phys.} 82, 451--497.
\newblock \doi{10.1103/RevModPhys.82.451}
\bibAnnoteFile{Sotiriou:2008rp}

\bibitem[{Sperhake et~al.(2017)Sperhake, Moore, Rosca, Agathos, Gerosa, and
  Ott}]{Sperhake:2017itk}
Sperhake, U., Moore, C.~J., Rosca, R., Agathos, M., Gerosa, D., and Ott, C.~D.
  (2017).
\newblock {Long-lived inverse chirp signals from core collapse in massive
  scalar-tensor gravity}.
\newblock \emph{Phys. Rev. Lett.} 119, 201103.
\newblock \doi{10.1103/PhysRevLett.119.201103}
\bibAnnoteFile{Sperhake:2017itk}

\bibitem[{Staykov et~al.(2014)Staykov, Doneva, Yazadjiev, and
  Kokkotas}]{Staykov:2014mwa}
Staykov, K.~V., Doneva, D.~D., Yazadjiev, S.~S., and Kokkotas, K.~D. (2014).
\newblock {Slowly rotating neutron and strange stars in $R^2$ gravity}.
\newblock \emph{JCAP} 1410, 006.
\newblock \doi{10.1088/1475-7516/2014/10/006}
\bibAnnoteFile{Staykov:2014mwa}

\bibitem[{Staykov et~al.(2015)Staykov, Doneva, Yazadjiev, and
  Kokkotas}]{Staykov:2015cfa}
Staykov, K.~V., Doneva, D.~D., Yazadjiev, S.~S., and Kokkotas, K.~D. (2015).
\newblock {Gravitational wave asteroseismology of neutron and strange stars in
  R$^2$ gravity}.
\newblock \emph{Phys. Rev.} D92, 043009.
\newblock \doi{10.1103/PhysRevD.92.043009}
\bibAnnoteFile{Staykov:2015cfa}

\bibitem[{Takami et~al.(2015)Takami, Rezzolla, and Baiotti}]{Takami:2014tva}
Takami, K., Rezzolla, L., and Baiotti, L. (2015).
\newblock {Spectral properties of the post-merger gravitational-wave signal
  from binary neutron stars}.
\newblock \emph{Phys. Rev.} D91, 064001.
\newblock \doi{10.1103/PhysRevD.91.064001}
\bibAnnoteFile{Takami:2014tva}

\bibitem[{{Thorne}(1969)}]{1969ApJ...158..997T}
{Thorne}, K.~S. (1969).
\newblock {Nonradial Pulsation of General-Relativistic Stellar Models.IV. The
  Weakfield Limit}.
\newblock \emph{Astrophys. J.} 158, 997.
\newblock \doi{10.1086/150259}
\bibAnnoteFile{1969ApJ...158..997T}

\bibitem[{Thorne(1980)}]{Thorne:1980ru}
Thorne, K.~S. (1980).
\newblock {Multipole Expansions of Gravitational Radiation}.
\newblock \emph{Rev. Mod. Phys.} 52, 299--339.
\newblock \doi{10.1103/RevModPhys.52.299}
\bibAnnoteFile{Thorne:1980ru}

\bibitem[{{Thorne} and {Campolattaro}(1967)}]{1967ApJ...149..591T}
{Thorne}, K.~S. and {Campolattaro}, A. (1967).
\newblock {Non-Radial Pulsation of General-Relativistic Stellar Models. I.
  Analytic Analysis for L $\ge$ 2}.
\newblock \emph{Astrophys. J.} 149, 591.
\newblock \doi{10.1086/149288}
\bibAnnoteFile{1967ApJ...149..591T}

\bibitem[{{Vaeth} and {Chanmugam}(1992)}]{1992A&A...260..250V}
{Vaeth}, H.~M. and {Chanmugam}, G. (1992).
\newblock {Radial oscillations of neutron stars and strange stars}.
\newblock \emph{Astronomy and Astrophysics} 260, 250--254
\bibAnnoteFile{1992A&A...260..250V}

\bibitem[{Völkel and Kokkotas(2019)}]{Volkel:2019gpq}
Völkel, S.~H. and Kokkotas, K.~D. (2019).
\newblock {On the Inverse Spectrum Problem of Neutron Stars}.
\newblock \emph{arXiv:1901.11262}
\bibAnnoteFile{Volkel:2019gpq}

\bibitem[{Wagoner(1970)}]{Wagoner:1970vr}
Wagoner, R.~V. (1970).
\newblock {Scalar tensor theory and gravitational waves}.
\newblock \emph{Phys. Rev. D} 1, 3209--3216.
\newblock \doi{10.1103/PhysRevD.1.3209}
\bibAnnoteFile{Wagoner:1970vr}

\bibitem[{Will(2006)}]{Will:2005va}
Will, C.~M. (2006).
\newblock {The Confrontation between general relativity and experiment}.
\newblock \emph{Living Rev. Rel.} 9, 3.
\newblock \doi{10.12942/lrr-2006-3}
\bibAnnoteFile{Will:2005va}

\bibitem[{Yagi and Yunes(2017)}]{Yagi:2016bkt}
Yagi, K. and Yunes, N. (2017).
\newblock {Approximate Universal Relations for Neutron Stars and Quark Stars}.
\newblock \emph{Phys. Rept.} 681, 1--72.
\newblock \doi{10.1016/j.physrep.2017.03.002}
\bibAnnoteFile{Yagi:2016bkt}

\bibitem[{Yazadjiev et~al.(2015)Yazadjiev, Doneva, and
  Kokkotas}]{Yazadjiev:2015zia}
Yazadjiev, S.~S., Doneva, D.~D., and Kokkotas, K.~D. (2015).
\newblock {Rapidly rotating neutron stars in R-squared gravity}.
\newblock \emph{Phys. Rev.} D91, 084018.
\newblock \doi{10.1103/PhysRevD.91.084018}
\bibAnnoteFile{Yazadjiev:2015zia}

\bibitem[{Yazadjiev et~al.(2017)Yazadjiev, Doneva, and
  Kokkotas}]{Yazadjiev:2017vpg}
Yazadjiev, S.~S., Doneva, D.~D., and Kokkotas, K.~D. (2017).
\newblock {Oscillation modes of rapidly rotating neutron stars in scalar-tensor
  theories of gravity}.
\newblock \emph{Phys. Rev.} D96, 064002.
\newblock \doi{10.1103/PhysRevD.96.064002}
\bibAnnoteFile{Yazadjiev:2017vpg}

\bibitem[{Yazadjiev et~al.(2014)Yazadjiev, Doneva, Kokkotas, and
  Staykov}]{Yazadjiev:2014cza}
Yazadjiev, S.~S., Doneva, D.~D., Kokkotas, K.~D., and Staykov, K.~V. (2014).
\newblock {Non-perturbative and self-consistent models of neutron stars in
  R-squared gravity}.
\newblock \emph{JCAP} 1406, 003.
\newblock \doi{10.1088/1475-7516/2014/06/003}
\bibAnnoteFile{Yazadjiev:2014cza}

\bibitem[{Zerilli(1970)}]{Zerilli:1970se}
Zerilli, F.~J. (1970).
\newblock {Effective potential for even parity Regge-Wheeler gravitational
  perturbation equations}.
\newblock \emph{Phys. Rev. Lett.} 24, 737--738.
\newblock \doi{10.1103/PhysRevLett.24.737}
\bibAnnoteFile{Zerilli:1970se}

\bibitem[{Özel and Freire(2016)}]{Ozel:2016oaf}
Özel, F. and Freire, P. (2016).
\newblock {Masses, Radii, and the Equation of State of Neutron Stars}.
\newblock \emph{Ann. Rev. Astron. Astrophys.} 54, 401--440.
\newblock \doi{10.1146/annurev-astro-081915-023322}
\bibAnnoteFile{Ozel:2016oaf}

\end{thebibliography}

\end{document}